\documentclass[aps,prl,reprint,superscriptaddress,showpacs]{revtex4-2}
\usepackage{graphicx}% Include figure files
\usepackage{dcolumn}% Align table columns on decimal point
\usepackage{bm}% bold math
 \usepackage{tensor}
 \usepackage{physics}
\usepackage{times}
\usepackage{siunitx}
\usepackage{textcomp }
\usepackage{xcolor }

\begin{document}

\title{Direct tomography of high-dimensional density matrices for general quantum states of photons}

\author{Yiyu Zhou}
\affiliation{The Institute of Optics, University of Rochester, Rochester, New York 14627, USA}
\author{Jiapeng Zhao}
\affiliation{The Institute of Optics, University of Rochester, Rochester, New York 14627, USA}
\author{Darrick Hay}
\affiliation{Department of Physics, University of South Florida, Tampa, Florida 33620, USA}
\author{Kendrick McGonagle}
\affiliation{Department of Physics, University of South Florida, Tampa, Florida 33620, USA}
\author{Robert W. Boyd}
\affiliation{The Institute of Optics, University of Rochester, Rochester, New York 14627, USA}
\affiliation{Department of Physics, University of Ottawa, Ottawa, Ontario K1N 6N5, Canada}
\author{Zhimin Shi}
\email{zshi.opt@gmail.com}
\affiliation{Department of Physics, University of South Florida, Tampa, Florida 33620, USA}

\date{\today}

\begin{abstract}
Quantum state tomography is the conventional method used to characterize density matrices for general quantum states. However, the data acquisition time generally scales linearly with the dimension of the Hilbert space, hindering the possibility of dynamic monitoring of a high-dimensional quantum system. Here, we demonstrate a direct tomography protocol to measure density matrices of photons in the position basis through the use of a polarization-resolving camera, where the dimension of density matrices can be as large as 580$\times$580 in our experiment. The use of the polarization-resolving camera enables parallel measurements in the position and polarization basis and as a result, the data acquisition time of our protocol does not increase with the dimension of the Hilbert space and is solely determined by the camera exposure time (on the order of 10 ms). Our method is potentially useful for the real-time monitoring of the dynamics of quantum states and paves the way for the development of high-dimensional, time-efficient quantum metrology techniques.
\end{abstract}
\maketitle
%Quantum mechanics plays a vital role in the development of many modern technologies for encoding, transporting, decoding, processing, and storing information\cite{MandelWolf_CoherentQuantumOpt, Kilin_PiO01:QuantaInfo, Dodonov_JOptB:NonclassicalStateReview}.

\textit{Introduction.}---The ability to characterize a quantum state is crucial in quantum technologies, both because it ensures that the desired quantum state has been generated and it can be used to determine the quantum state after interacting with a system. Quantum state tomography is an established approach to reconstruct a general quantum state (either pure or mixed) through a series of projective measurements performed on identically prepared states \cite{White_PRA01:CharacterizeQuanInfo, Itatani_Nat04:TomographicImaging, Resch_PRL05:ThreePhotonStateTomography, Soderholm_NJP12:PolaTomography, Sych_PRA12:InfoCompleteContVariable,PhysRevLett.87.050402, Beck_PRL00:QuanTomoDetArray,paris2004quantum, Beck_PRL01:QuanMeasureDetArray, Dawes_PRA03:quantumTomowDetArray, Smith_OL05:Measure_SglPhoton, PhysRevLett.70.1244, RevModPhys.81.299}. Recently, the concept of direct measurement \cite{Lundeen_Nat11:DirectMeasure} has been established, which can directly be used to read out the complex-valued amplitudes of a pure quantum state through a proper sequence of weak and strong measurements \cite{Aharonov_PRL88:WeakMeasurementSpin, Duck_PRD89:WeakMeasurement, Ritchie_PRL91:RealizationWeakMeasurement, Johansen_PRL04:WeakMeasurementArbitraryProbeStates, Hosten_Science04:SpinHallEff_weakMeasure, Solli_PRL04:FLSL_GeneralizedWeakValue,DixonPRL09:WeakDeflection,Feizpour_PRL11:WeakSinglePhotonNonlinearity, Kocsis_Science12:EMCCD_entanglement, Dressel_13ti:WeakReview, PhysRevLett.112.070405, turek2015post, ren2019efficient, ogawa2019framework, vallone2016strong, zou2015direct, PhysRevA.101.012119}. The elimination of the complicated post-processing procedure of state reconstruction is one of the main advantages of direct measurement methods, allowing it to serve as an alternative metrology technique that may greatly reduce experimental complexity.

The concept of direct measurement is quickly being extended to the characterization of various quantum systems \cite{Thekkadath2016Direct, Lundeen_PRL12:WeakMeasureGeneral, Wu_SciRep:StateTomography, Saivail_NatPhon13:PolarizationDirectMeasure, Mirhosseini_PRL14:CompressiveDM,Malik_13tj:DirectMeasureOAM, Mirhosseini2016Wigner}. Nonetheless, one remaining challenge in quantum-state metrology is the limited characterization speed and efficiency for high-dimensional quantum states. Most demonstrated techniques, including direct measurement methods, involve either a slow scanning process or a complicated post-processing procedure, where the characterization time scales unfavorably with the dimension of the quantum system. As a result, almost all quantum metrology demonstrations to date have been carried out under stable laboratory conditions, and the measurement of a high-dimensional quantum state can take as long as several hours. Compressive sensing has been implemented for the tomography of an $N$-dimensional pure state in the spatial domain with $N=19,200$, which still requires $\approx${}$0.25N$ measurements \cite{Mirhosseini_PRL14:CompressiveDM}{}. Direct measurement of the density matrix in the high-dimensional orbital-angular-momentum (OAM) basis has also been reported \cite{Mirhosseini2016Wigner, Malik_13tj:DirectMeasureOAM}{}. However, these methods use single-pixel detectors for data collection and require performing a series of measurements via scanning for the reconstruction of high-dimensional quantum states. In general, since the number of measurements scales linearly with the dimension of the Hilbert space, the data acquisition time inevitably increases for high-dimensional quantum states, hindering the possibility of real-time monitoring of dynamic quantum systems. While the recently proposed auxiliary Hilbert space tomography \cite{liu2019efficient}{} can reduce the measurement complexity for density matrix characterization, this method is only applicable to OAM states and thus exhibits a limited range of application. This is because it is more desirable to characterize the density matrix in the position basis, which is analogous to the mutual coherence function in classical optics, of which Michelson stellar interferometry \cite{mandel1995optical}{} is a practical application. However, the conventional method observes the interference visibility between two apertures to measure the coherence between two points, and the positions of two apertures have to be scanned to obtain the complete density matrix \cite{hradil2010quantum}{}, which is extremely time consuming. In the following, we introduce a scan-free direct tomography protocol that can measure the complex-valued high-dimensional density matrix for mixed photon states in the position basis by using a polarization-resolving camera. The data acquisition time does not increase with the dimension of the Hilbert space, and the maximum dimension allowed by our protocol is only limited by the pixel count of the detector array.

% \begin{equation}
% \begin{aligned}
% \hat{\rho}_0 = \sum_{k} p_k \ket{\psi_k}\bra{\psi_k},
% \end{aligned}
% \end{equation}

\textit{Direct tomography protocol.}---The density matrix can be represented as an incoherent mixture of pure states, which can be expressed as $ \hat{\rho}_0 = \sum_{k} p_k \ket{\psi_k}\bra{\psi_k}$, where $\ket{\psi_k}$ is the pure quantum state normalized as $\bra{\psi_k}\ket{\psi_k}=1$, and $p_k$ is the probability coefficient normalized as $\sum_{k} p_k=1$. The element of a density matrix in the position basis can be computed as
\begin{equation}
\begin{aligned}
\rho_0(x_1,y_1,x_2,y_2) &= \bra{x_1,y_1} \hat{\rho}_0 \ket{x_2,y_2} \\
&= \sum_{k} p_k \bra{x_1,y_1}\ket{\psi_k}\bra{\psi_k}\ket{x_2,y_2} \\
& = \sum_{k} p_k \psi_k(x_1,y_1)\psi^*_k(x_2,y_2),
\end{aligned}
\label{eq:rho0element}
\end{equation}
where $\ket{x,y}$ denotes the position eigenstate located at $(x,y)$. In our experiment, we assume that the transverse profile of the quantum state has only a one-dimensional (1D) variation along the $x$ axis and is invariant along the $y$ axis. Therefore, we have $\psi_k(x_1,y_1)=\psi_k(x_1)$ is independent of $y$, and the density matrix element can be simplified as $\rho_0(x_1,y_1,x_2,y_2) =\rho_{0}(x_1,x_2)$, and
\begin{figure}[t]
\center
\includegraphics[width= \linewidth]{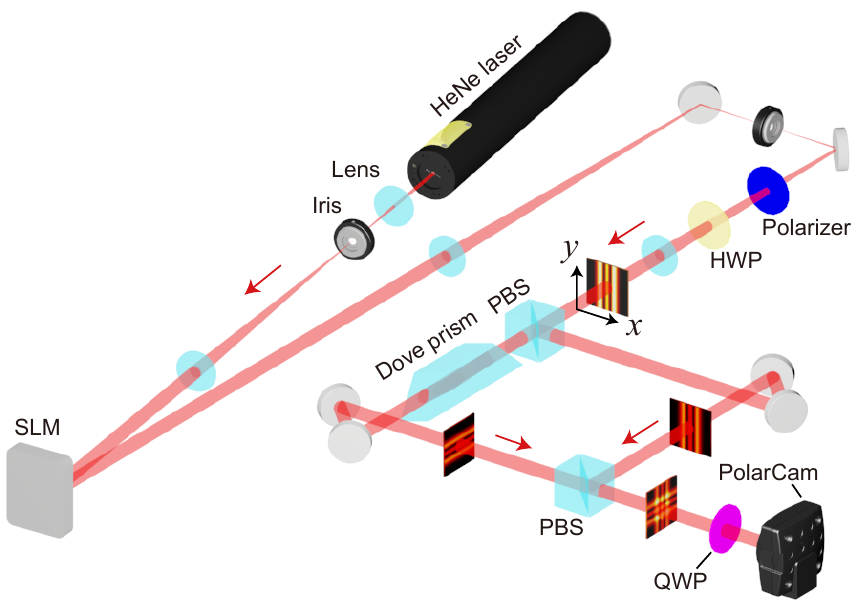}
\caption{The schematic of the experimental setup. A Dove prism is used to rotate the horizontally polarized beam by 90\textdegree{}. A 1D Hermite-Gauss state HG$_2(x)$ is used as an example to visualize the beam rotation. SLM: spatial light modulator. HWP: half-wave plate. PBS: polarizing beamsplitter. QWP: quarter-wave plate. PolarCam: polarization-resolving camera.}
\label{fig:setup}
\end{figure}
\begin{equation}
\begin{aligned}
\rho_{0}(x_1,x_2)= \sum_{k} p_k \psi_k(x_1)\psi^*_k(x_2),
\label{eq:element}
\end{aligned}
\end{equation}
which is the quantity to be measured. It is worth noting that Eq.~(\ref{eq:element}) is reminiscent of the mutual coherence function in classical optics \cite{mandel1995optical, hradil2010quantum}. We use the polarization as the the pointer state \cite{Thekkadath2016Direct}{} which is prepared in the diagonal polarization state $\ket{{\rm{D}}}=(\ket{{\rm{H}}}+\ket{\rm{V}})/\sqrt{2}$, where $\ket{{\rm{H}}}$ and $\ket{\rm{V}}$ denote the horizontal and vertical polarization state, respectively. Therefore, the full initial density matrix can be written as $\hat{\rho}_i = \hat{\rho}_0 \otimes \ket{{\rm{D}}}\bra{{\rm{D}}}$. Our direct tomography protocol entails introducing a 90\textdegree{} beam rotation for the horizontally polarized beam while leaving the vertically polarized beam unchanged. This polarization-sensitive beam rotation can be described by a unitary transformation as \cite{Mirhosseini2016Wigner}{}
\begin{equation}
\begin{aligned}
\hat{{\rm{U}}}=\hat{{\rm{T}}}(\pi/2) \otimes \ket{{\rm{H}}}\bra{{\rm{H}}}+ \hat{{\rm{T}}}(0)\otimes \ket{{\rm{V}}}\bra{{\rm{V}}},
\end{aligned}
\label{eq:unitary}
\end{equation}
where $\hat{{\rm{T}}}(\theta)=\exp (-i \theta \hat{\ell} )$ is the rotation operator, $\hat{{\rm{\ell}}}$ is the orbital angular momentum operator about the optical axis, and the effect of rotation operator on the position eigenstate can be written as $\hat{{\rm{T}}}(\theta)\ket{x,y}=\ket{x\cos \theta+y \sin \theta, -x\sin \theta + y \cos \theta}$. The final density matrix after this unitary transformation can be represented as $\hat{\rho}_f= \hat{{\rm{U}}} \hat{\rho}_i \hat{{\rm{U}}}^{\dagger}$. The projective measurements \cite{Zhimin2015Scan, Single2019Zhu} we propose to perform can be represented by the following projectors:
\begin{equation}
\begin{aligned}
\hat{\bm{\pi}}_{\rm{D}} &= \ket{x,y}\bra{x,y} \otimes \ket{{\rm{D}}}\bra{{\rm{D}}}, \\
\hat{\bm{\pi}}_{\rm{A}} &= \ket{x,y}\bra{x,y} \otimes \ket{{\rm{A}}}\bra{{\rm{A}}}, \\
\hat{\bm{\pi}}_{\rm{R}} &= \ket{x,y}\bra{x,y} \otimes \ket{{\rm{R}}}\bra{{\rm{R}}},\\
\hat{\bm{\pi}}_{\rm{L}} &= \ket{x,y}\bra{x,y} \otimes \ket{{\rm{L}}}\bra{{\rm{L}}},
\end{aligned}
\label{eq:projectors}
\end{equation}
where $\ket{{\rm{A}}}=(\ket{{\rm{H}}}-\ket{{\rm{V}}})/\sqrt{2}$ is the anti-diagonal polarization state, $\ket{{\rm{L}}}=(\ket{{\rm{H}}}+i\ket{{\rm{V}}})/\sqrt{2}$ is the left-handed circular polarization state, and $\ket{{\rm{R}}}=(\ket{{\rm{H}}}-i\ket{{\rm{V}}})/\sqrt{2}$ is the right-handed circular polarization state. Therefore, the expectation value of these projectors are found to be
\begin{equation}
\begin{aligned}
\Gamma_{\rm{D}}(x,y) &= {\rm{Tr}}[\hat{\bm{\pi}}_{\rm{D}} \hat{{\rm{U}}} \hat{\rho}_i \hat{{\rm{U}}}^{\dagger} ] \\
&=\frac{1}{4} \left( \rho_0(-y,-y)+\rho_0(x,x)+2{\rm{Re}}[\rho_0(-y,x)] \right), \\
\Gamma_{\rm{A}}(x,y) &= {\rm{Tr}}[\hat{\bm{\pi}}_{\rm{A}} \hat{{\rm{U}}} \hat{\rho}_i \hat{{\rm{U}}}^{\dagger} ] \\
&=\frac{1}{4} \left( \rho_0(-y,-y)+\rho_0(x,x)-2{\rm{Re}}[\rho_0(-y,x)] \right), \\
\Gamma_{\rm{R}}(x,y) &= {\rm{Tr}}[\hat{\bm{\pi}}_{\rm{R}} \hat{{\rm{U}}} \hat{\rho}_i \hat{{\rm{U}}}^{\dagger} ] \\
&=\frac{1}{4} \left( \rho_0(-y,-y)+\rho_0(x,x)+2{\rm{Im}}[\rho_0(-y,x)] \right), \\
\Gamma_{\rm{L}}(x,y) &= {\rm{Tr}}[\hat{\bm{\pi}}_{\rm{L}} \hat{{\rm{U}}} \hat{\rho}_i \hat{{\rm{U}}}^{\dagger} ] \\
&=\frac{1}{4} \left( \rho_0(-y,-y)+\rho_0(x,x)-2{\rm{Im}}[\rho_0(-y,x)] \right).
\end{aligned}
\end{equation}

Using the above equations, the density matrix can be experimentally reconstructed as
\begin{equation}
\begin{aligned}
\rho^{{\rm{exp}}}_0(x_1,x_2) &= \Gamma_{\rm{D}}(x_2,-x_1)- \Gamma_{\rm{A}}(x_2,-x_1) \\
&+i(\Gamma_{\rm{R}}(x_2,-x_1)- \Gamma_{\rm{L}}(x_2,-x_1)).
\label{eq:DMrecon}
\end{aligned}
\end{equation}
It can be seen that the density matrix can be directly reconstructed without using any complicated algorithm. In addition to the reconstruction of the density matrix $\hat{\rho}^{{\rm{exp}}}_0$, it is also desirable to be able to reconstruct the pure states $\ket{\psi_k}$. In order to reconstruct the pure states, we use singular value decomposition \cite{nielsenchuang2010}{}. The reconstruction can be unique if the pure states are mutually orthogonal. For a square and Hermitian density matrix $\hat{\rho}^{{\rm{exp}}}_0$, it can always be decomposed as \cite{trefethen1997numerical}{}
\begin{equation}
\begin{aligned}
\hat{\rho}^{{\rm{exp}}}_0=\hat{{\rm{M}}} \hat{{\rm{S}}}\hat{{\rm{M}}}^{\dagger},
\end{aligned}
\end{equation}
where $\hat{{\rm{M}}}$ is a unitary matrix, and $\hat{{\rm{S}}}$ is a real-valued diagonal matrix whose diagonal elements ${\rm{S}}_{kk}$ are the singular values of $\hat{\rho}^{{\rm{exp}}}_0$. It can be readily seen that
\begin{equation}
\begin{aligned}
\bra{x_1} \hat{\rho}^{{\rm{exp}}}_0 \ket{x_2}=\sum_{k} {\rm{S}}_{kk} \bra{x_1} \hat{{\rm{M}}} \ket{k} \bra{k} \hat{{\rm{M}}}^{\dagger} \ket{x_2}.
\end{aligned}
\label{eq:svd2}
\end{equation}
Comparing Eq.~(\ref{eq:svd2}) with Eq.~(\ref{eq:rho0element}), one can find that
\begin{equation}
\begin{aligned}
p_k^{{\rm{exp}}}&={\rm{S}}_{kk}, \\
\bra{x}\ket{\psi_k^{{\rm{exp}}}}&=\bra{x} \hat{{\rm{M}}} \ket{k}.
\end{aligned}
\label{eq:svdrecon}
\end{equation}
As one can see, singular value decomposition can be used as a tool to decompose a density matrix into an incoherent mixture of pure states, which can be efficiently implemented by established numerical algorithms \cite{Klema1980singular}. It is worth noting that the singular value decomposition discussed here is reminiscent of the coherent mode decomposition in optical coherence theory \cite{Wolf82, WOLF19813}{}.

\begin{figure}[t]
\center
\includegraphics[width= \linewidth]{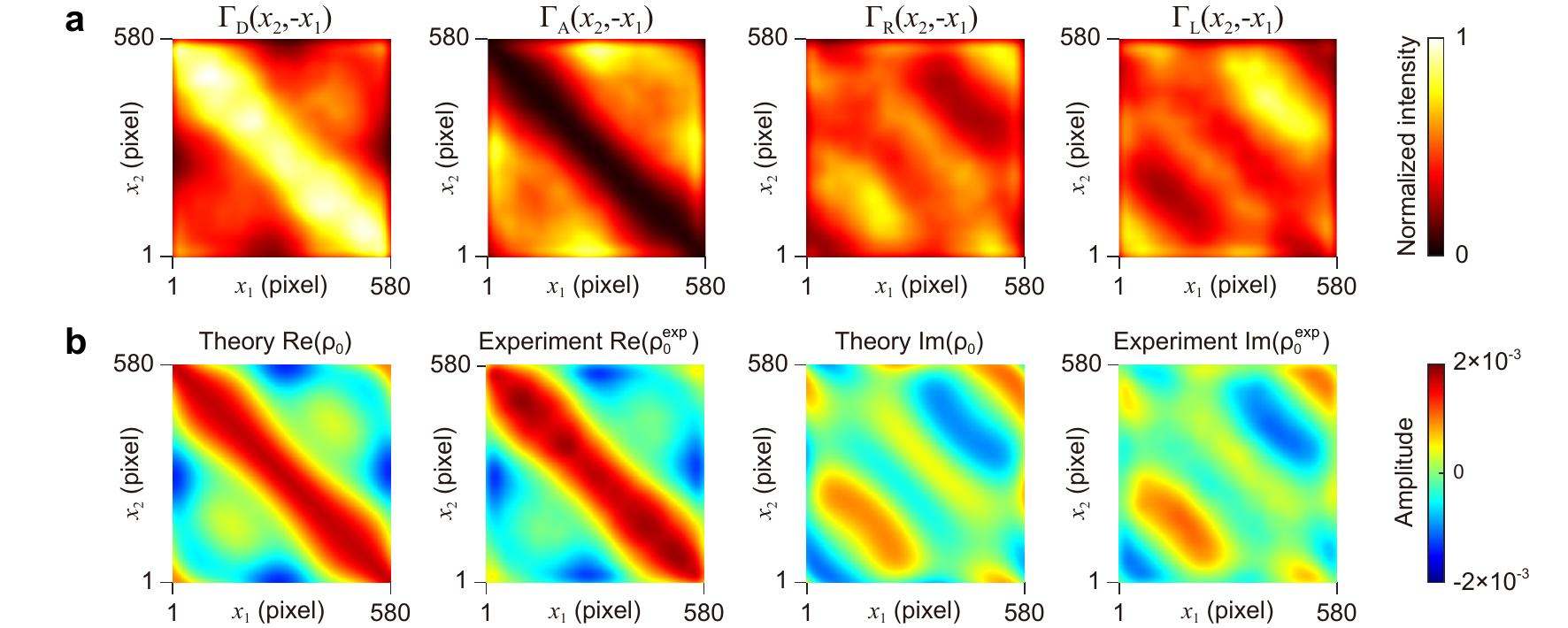}
\caption{Experimental results for the phase-only states. (a) The images acquired by the PolarCam. (b) The real and imaginary part of the reconstructed density matrix. The trace distance between the theoretical density matrix and the experimentally measured density matrix is 14.2\%$\pm$0.3\%.}
\label{fig:DM_Phase}
\end{figure}

\textit{Experiment.}---The experimental setup to implement the direct tomography protocol is shown in Fig.~\ref{fig:setup}. A 633~nm HeNe laser with an optical power of 3 mW is used as the source of photons. The light beam is spatially filtered and attenuated before it illuminates a spatial light modulator (SLM, Pluto 2 VIS-020, Holoeye). A series of computer-generated phase-only holograms \cite{Arrizon07}{} is displayed onto the SLM to generate the quantum states of interest. Mixed states can be generated by switching the hologram on the SLM and by incoherently mixing the intensity images acquired by the camera \cite{Rodenburg2014Experimental}{}. An iris is used to pass the first diffraction order of light coming off the SLM while blocking all other diffraction orders. A polarizer and a half-wave plate (HWP) are used to generate the diagonal polarization state $\ket{{\rm{D}}}$. To implement the unitary transformation $\hat{{\rm{U}}}$ [cf. Eq.~(\ref{eq:unitary})], a polarizing beamsplitter (PBS) is used to separate the horizontally and vertically polarized beam. A Dove prism is applied to geometrically rotate the horizontally polarized beam by 90\textdegree{}. A second PBS is used to recombine the two beams and thus completes the implementation of $\hat{{\rm{U}}}$. A 45\textdegree{}-oriented quarter-wave plate (QWP) and a polarization-resolving camera (PolarCam, BFS-U3-51S5P-C, FLIR) are used to perform all the required projective measurements in a single shot. The PolarCam has micro-sized polarizers (oriented to 0\textdegree{}, 45\textdegree{}, 90\textdegree{}, and 135\textdegree{}, respectively) deposited on the camera sensors and thus allows for the detection of four different polarization states simultaneously. The camera exposure time is approximately 10 milliseconds depending on the intensity of the generated states. The QWP and the PolarCam jointly enable the projective measurements proposed in Eq.~(\ref{eq:projectors}). The image on the camera has a size of 580$\times$580 pixels, and thus the dimensionality of the quantum states in our experiment is $N=580$. The pixel size of the camera is 3.45~$\mu$m.

\begin{figure}[t]
\center
\includegraphics[width= \linewidth]{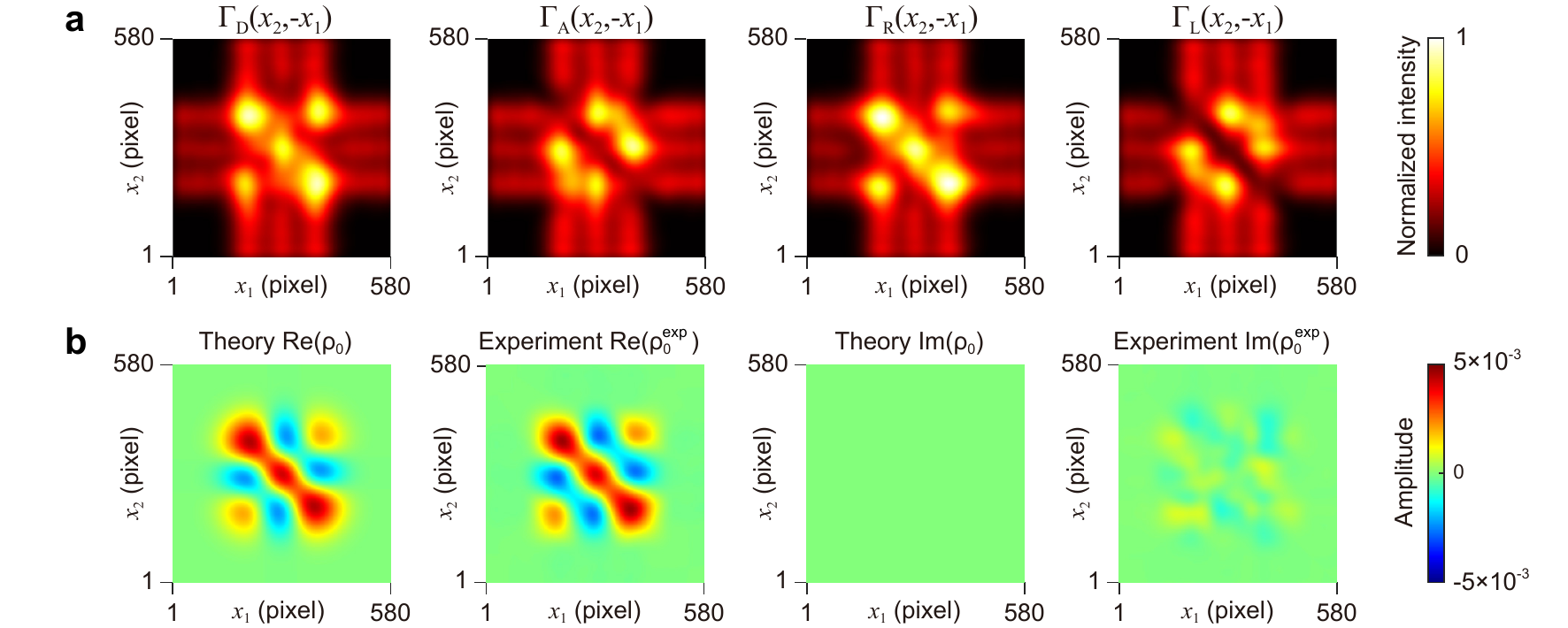}
\caption{Experimental results for the HG states. (a) The images acquired by the PolarCam. (b) The real and imaginary part of the reconstructed density matrix. The trace distance between the theoretical density matrix and the experimentally measured density matrix is 19.0\%$\pm$0.3\%.}
\label{fig:DM_HG}
\end{figure}

In our experiment, we prepare a mixed state consisting of three mutually orthogonal pure states $\ket{\psi_k}$ with $k=1,2,3$. More specifically, as our first demonstration, the pure states used to construct the density matrix are
\begin{equation}
\begin{aligned}
p_1=0.21, \quad \bra{x_{}}\ket{\psi_1}&=e^{i1.04 \pi x_{}/a} , \\
p_2=0.30, \quad \bra{x_{}}\ket{\psi_2}&=e^{i\pi [-8.42(x_{}/a)^3+4.04(x_{}/a)]}, \\
p_3=0.49, \quad \bra{x_{}}\ket{\psi_3}&=e^{i\pi [-17.6(x_{}/a)^5-x_{}/a]},
\end{aligned}
\end{equation}
where $-a/2 < x_{} \leq a/2$ is the discretized position, and $a=2$~mm is the size of the beam. These states are referred to the phase-only quantum states henceforth.

As another test of our protocol, we use the 1D Hermite-Gauss (HG) states to construct the mixed state:
\begin{equation}
\begin{aligned}
{\rm{HG}}_m(x_{}) =& \left(\frac{2}{\pi w_0^2} \right)^{\frac{1}{4}} \frac{1}{\sqrt{2^m m!}} \\
& \times H_m\left( \frac{\sqrt{2}x_{}}{w_0} \right) \exp(\frac{-x_{}^2}{w_0^2}),
\end{aligned}
\label{eq:1DHGmodes}
\end{equation}
where $H_m(\cdot)$ is the Hermite polynomial of order $m$ \cite{svelto2010principles}{}, and $w_0=0.15a$ is the beam waist radius. The HG states used in the experiment are
\begin{equation}
\begin{aligned}
p_1=0.22, \quad \bra{x_{}}\ket{\psi_1}&={\rm{HG}}_0(x_{}), \\
p_2=0.33, \quad \bra{x_{}}\ket{\psi_2}&={\rm{HG}}_1(x_{}),\\
p_3=0.45, \quad\bra{x_{}}\ket{\psi_3}&={\rm{HG}}_2(x_{}) .
\end{aligned}
\label{eq:HGStates}
\end{equation}

\begin{figure}[t]
\center
\includegraphics[width= 1\linewidth]{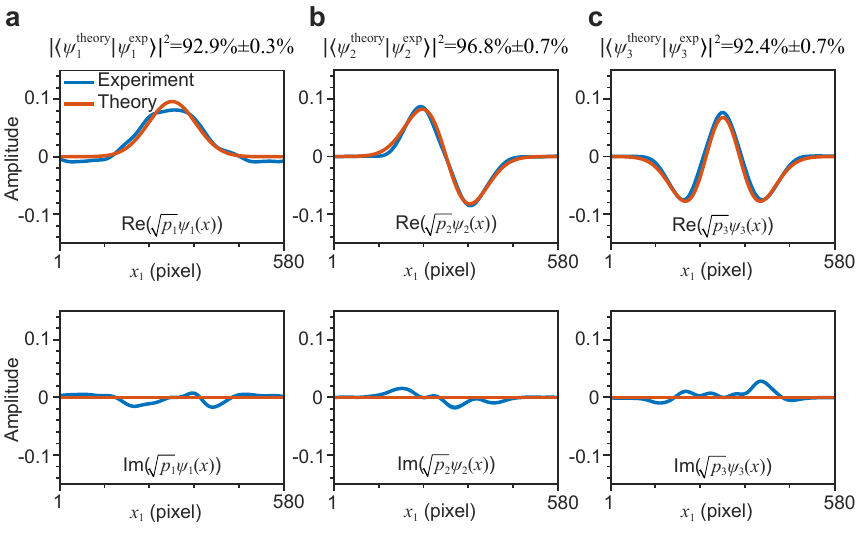}
\caption{The reconstructed HG quantum state for (a) $\ket{\psi_1}$, (b) $\ket{\psi_2}$, and (c) $\ket{\psi_3}$, respectively. The real (imaginary) part is shown in the left (right) panel. The standard deviation of the experimental data is denoted by the line width, which is generally too small to be visible. The fidelity of each state reconstructed is shown at the top of each corresponding subfigure.}
\label{fig:Mode_HG}
\end{figure}

The images acquired by the PolarCam for the phase-only states are shown in Fig.~\ref{fig:DM_Phase}(a). We apply a digital low-pass Gaussian spatial filter to process these images in order to remove the undesirable fringes caused by dusts and glass film interference \cite{GaussFilter}. The density matrix can be directly reconstructed based on these data by using Eq.~(\ref{eq:DMrecon}). Due to the experimental errors (e.g., misalignments, noises, imperfect mode generation fidelity, etc.), the experimentally reconstructed density matrix $\hat{\rho}_{0}^{\rm{exp}}$ may not be strictly Hermitian. Hence, we implement $\hat{\rho}_{0}^{\rm{exp}} \rightarrow (\hat{\rho}_{0}^{\rm{exp}} +\hat{\rho}_{0}^{\rm{exp} \dagger} )/2$ to guarantee the Hermiticity of the density matrix, and the results are shown in Fig.~\ref{fig:DM_Phase}(b). To quantify the accuracy of our protocol, we calculate the trace distance between the ideal density matrix $\hat{\rho_0}$ and the experimentally measured density matrix $\hat{\rho}_0^{\rm{exp}}$ as follows \cite{nielsenchuang2010}{}:
\begin{equation}
\begin{aligned}
\text{Trace distance}=\frac{1}{2} \left| {\rm{Tr}} [\sqrt{ (\hat{\rho_0}-\hat{\rho}_0^{\rm{exp}})(\hat{\rho_0}-\hat{\rho}_0^{\rm{exp}})^{\dagger} }] \right|,
\end{aligned}
\label{eq:tracedistance}
\end{equation}
\begin{figure}[t]
\center
\includegraphics[width=1 \linewidth]{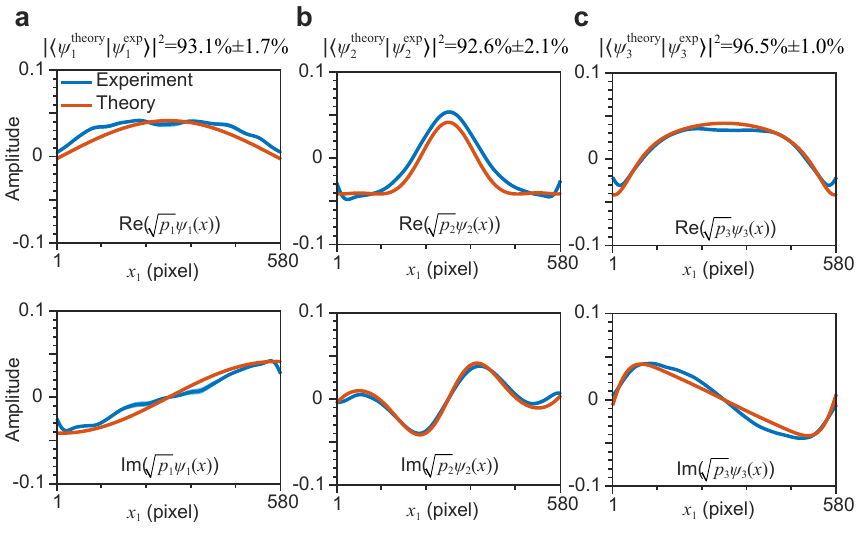}
\caption{The reconstructed phase-only quantum state for (a) $\ket{\psi_1}$, (b) $\ket{\psi_2}$, and (c) $\ket{\psi_3}$, respectively. The real (imaginary) part is shown in the left (right) panel. The standard deviation of the experimental data is denoted by the line width, which is generally too small to be visible. The fidelity of each reconstructed state is shown at the top of each corresponding subfigure.}
\label{fig:Mode_Phase}
\end{figure}
and the trace distance for the phase-only states is calculated to be 14.2\%$\pm$0.3\%. It should be noted that a lower trace distance indicates a higher measurement fidelity of our measurement protocol. This is because the trace distance quantifies the maximum possible probability of distinguishing the quantum states described by two density matrices, and thus the trance distance between two exactly identical states is zero. The experimental results for the HG states are presented in Fig.~\ref{fig:DM_HG}, and the corresponding trace distance is measured to be 19.0\%$\pm$0.3\%. We also numerically perform the singular value decomposition for the experimentally measured density matrix. The reconstructed phase-only states are shown in Fig.~\ref{fig:Mode_Phase}, and the reconstructed HG states are shown in Fig.~\ref{fig:Mode_HG}. For each experimentally reconstructed quantum state $\ket{\psi_k^{\rm{exp}}}$, we compute its fidelity as $\left|\bra{\psi_k^{\rm{theory}}}\ket{\psi_k^{\rm{exp}}}\right|^2$. The fidelity for each reconstructed state is shown at the top of each corresponding subfigure. It can be seen that the fidelity of state is always higher than 90\%. In our experiment, we attribute the nonzero trace distance primarily to the imperfect spatial mode generation and the misalignment of the polarization-sensitive beam rotator. As a consequence, the reconstructed density matrix might be unphysical due to the possible lack of Hermiticity and positive semi-definiteness \cite{Measurement2001James}{}. However, we notice that the standard maximum-likelihood-estimation-based routine for recovering a physical density matrix \cite{Measurement2001James}{} is not readily applicable to our experiment, because it requires the minimization of a likelihood function with $N^2=336,400$ independent parameters. This task can potentially be accomplished by using machine learning algorithms \cite{torlai2018neural}{} and is subject to future study. In our experiment, we assume the transverse profile of the field has only a 1D variation along the $x$ axis [see Eq.~(\ref{eq:element})]. Although our protocol cannot be directly applied to a general two-dimensional (2D) spatial field, it is possible to reshape a finite-sized 2D field into a 1D field \cite{berkhout2010efficient}{} to further generalize our approach. A potential realization of the 2D-to-1D beam reshaping is discussed in Supplemental Material \cite{Supplement}{}, which includes Refs. \cite{park2017large, Zhimin2015Scan, mandel1995optical, Charbon_PTRSLA14:SinglePhotonImagingSPADReview, Gariepy_NatComm15:SinglePhotonFlightImaging, Dawes_PRA03:quantumTomowDetArray, Kocsis_Science12:EMCCD_entanglement, Edgar_NatComm12:EMCCD_entanglement, Lemos_Nat14:QIm_w_undetectedPhotons, Tasca_OpEx13:GI_ICCD, Aspden_NJP13:EPRGI_using_ICCD, Fickler_SciRep:RealTimeImagingEntanglement, Machulka_OpEx14:SpatialPropofTwinbeamCorrelation, Morris_NatComm15:ImagingSinglePhotons, Chrapkiewicz_NatPhoton16:ImagingSinglePhotons}.

Although many quantum techniques use single-pixel detectors, advances in detector development have led to many options for the use of high-performance detector arrays, such as SPAD arrays \cite{Charbon_PTRSLA14:SinglePhotonImagingSPADReview, Gariepy_NatComm15:SinglePhotonFlightImaging}, cooled CCD cameras \cite{Dawes_PRA03:quantumTomowDetArray, Kocsis_Science12:EMCCD_entanglement}, electron-multiplying CCD cameras \cite{Edgar_NatComm12:EMCCD_entanglement, Lemos_Nat14:QIm_w_undetectedPhotons} and intensified CCD cameras \cite{Tasca_OpEx13:GI_ICCD, Aspden_NJP13:EPRGI_using_ICCD, Fickler_SciRep:RealTimeImagingEntanglement, Machulka_OpEx14:SpatialPropofTwinbeamCorrelation, Morris_NatComm15:ImagingSinglePhotons, Chrapkiewicz_NatPhoton16:ImagingSinglePhotons}. Comparing with raster scanning techniques using a single-pixel detector, the parallel measurement via a $M$-pixel detector array is generally $M$ times faster, which can be used to apply our method to quantum applications at the single-photon level. It is worth mentioning that due to the photon loss induced by the polarizers in the PolarCam, the photon efficiency of our method is suboptimal. However, the PolarCam can in principle be replaced by polarizing beamsplitters and a regular camera to eliminate the photon loss \cite{Zhimin2015Scan}. In contrast, the standard raster scanning technique requires the use of two scanning apertures, in which the photon efficiency drops by a factor of $N$ for a $N$-dimensional photon state due to the aperture postselection loss. Meanwhile, the scanning technique also would increase the measuring time by a factor of $N^2$. Therefore, our method can significantly outperform the standard raster scanning technique in terms of both measurement speed and photon efficiency (see Supplemental Material \cite{Supplement} for details).

\textit{Conclusion.}---In this work, we demonstrated a direct tomography protocol that can efficiently characterize a high-dimensional density matrix in the position basis for general quantum states, where the data acquisition time is independent of the dimension of the Hilbert space. Two different mixed states were prepared and characterized with a high fidelity in our demonstration. Singular value decomposition was implemented to reconstruct the pure states that constitute the prepared mixed state, which can potentially be useful for the analysis of spatially incoherent fields. We anticipate that our protocol can inspire the development of high-dimensional, time-efficient quantum metrology techniques and can be used as a powerful tool for the experimental study of the spatial mutual coherence function of optical fields, which plays an important role in Michelson stellar interferometry, the super-resolution imaging \cite{larson2018resurgence} and optical coherence theory \cite{mandel1995optical}{}.

\begin{acknowledgments}
This work is supported by the U.S. Office of Naval Research (N00014-17-1-2443, N00014-20-1-2558). In addition, R.W.B. acknowledges support from Canada Research Chairs Program and Natural Sciences and Engineering Research Council of Canada.
\end{acknowledgments}

%\bibliographystyle{apsrev4-2}
%\bibliography{DensityMatrix}

\begin{thebibliography}{68}%
\makeatletter
\providecommand \@ifxundefined [1]{%
 \@ifx{#1\undefined}
}%
\providecommand \@ifnum [1]{%
 \ifnum #1\expandafter \@firstoftwo
 \else \expandafter \@secondoftwo
 \fi
}%
\providecommand \@ifx [1]{%
 \ifx #1\expandafter \@firstoftwo
 \else \expandafter \@secondoftwo
 \fi
}%
\providecommand \natexlab [1]{#1}%
\providecommand \enquote  [1]{``#1''}%
\providecommand \bibnamefont  [1]{#1}%
\providecommand \bibfnamefont [1]{#1}%
\providecommand \citenamefont [1]{#1}%
\providecommand \href@noop [0]{\@secondoftwo}%
\providecommand \href [0]{\begingroup \@sanitize@url \@href}%
\providecommand \@href[1]{\@@startlink{#1}\@@href}%
\providecommand \@@href[1]{\endgroup#1\@@endlink}%
\providecommand \@sanitize@url [0]{\catcode `\\12\catcode `\$12\catcode
  `\&12\catcode `\#12\catcode `\^12\catcode `\_12\catcode `\%12\relax}%
\providecommand \@@startlink[1]{}%
\providecommand \@@endlink[0]{}%
\providecommand \url  [0]{\begingroup\@sanitize@url \@url }%
\providecommand \@url [1]{\endgroup\@href {#1}{\urlprefix }}%
\providecommand \urlprefix  [0]{URL }%
\providecommand \Eprint [0]{\href }%
\providecommand \doibase [0]{https://doi.org/}%
\providecommand \selectlanguage [0]{\@gobble}%
\providecommand \bibinfo  [0]{\@secondoftwo}%
\providecommand \bibfield  [0]{\@secondoftwo}%
\providecommand \translation [1]{[#1]}%
\providecommand \BibitemOpen [0]{}%
\providecommand \bibitemStop [0]{}%
\providecommand \bibitemNoStop [0]{.\EOS\space}%
\providecommand \EOS [0]{\spacefactor3000\relax}%
\providecommand \BibitemShut  [1]{\csname bibitem#1\endcsname}%
\let\auto@bib@innerbib\@empty
%</preamble>
\bibitem [{\citenamefont {White}\ \emph {et~al.}(2001)\citenamefont {White},
  \citenamefont {James}, \citenamefont {Munro},\ and\ \citenamefont
  {Kwiat}}]{White_PRA01:CharacterizeQuanInfo}%
  \BibitemOpen
  \bibfield  {author} {\bibinfo {author} {\bibfnamefont {A.~G.}\ \bibnamefont
  {White}}, \bibinfo {author} {\bibfnamefont {D.~F.~V.}\ \bibnamefont {James}},
  \bibinfo {author} {\bibfnamefont {W.~J.}\ \bibnamefont {Munro}},\ and\
  \bibinfo {author} {\bibfnamefont {P.~G.}\ \bibnamefont {Kwiat}},\ }\href
  {https://doi.org/10.1103/PhysRevA.65.012301} {\bibfield  {journal} {\bibinfo
  {journal} {Phys. Rev. A}\ }\textbf {\bibinfo {volume} {65}},\ \bibinfo
  {pages} {012301} (\bibinfo {year} {2001})}\BibitemShut {NoStop}%
\bibitem [{\citenamefont {Itatani}\ \emph {et~al.}(2004)\citenamefont
  {Itatani}, \citenamefont {Levesque}, \citenamefont {Zeidler}, \citenamefont
  {Niikura}, \citenamefont {P\'{e}pin}, \citenamefont {Kieffer}, \citenamefont
  {Corkum},\ and\ \citenamefont
  {Villeneuve}}]{Itatani_Nat04:TomographicImaging}%
  \BibitemOpen
  \bibfield  {author} {\bibinfo {author} {\bibfnamefont {J.}~\bibnamefont
  {Itatani}}, \bibinfo {author} {\bibfnamefont {J.}~\bibnamefont {Levesque}},
  \bibinfo {author} {\bibfnamefont {D.}~\bibnamefont {Zeidler}}, \bibinfo
  {author} {\bibfnamefont {H.}~\bibnamefont {Niikura}}, \bibinfo {author}
  {\bibfnamefont {H.}~\bibnamefont {P\'{e}pin}}, \bibinfo {author}
  {\bibfnamefont {J.~C.}\ \bibnamefont {Kieffer}}, \bibinfo {author}
  {\bibfnamefont {P.~B.}\ \bibnamefont {Corkum}},\ and\ \bibinfo {author}
  {\bibfnamefont {D.~M.}\ \bibnamefont {Villeneuve}},\ }\href@noop {}
  {\bibfield  {journal} {\bibinfo  {journal} {Nature}\ }\textbf {\bibinfo
  {volume} {432}},\ \bibinfo {pages} {867} (\bibinfo {year}
  {2004})}\BibitemShut {NoStop}%
\bibitem [{\citenamefont {Resch}\ \emph {et~al.}(2005)\citenamefont {Resch},
  \citenamefont {Walther},\ and\ \citenamefont
  {Zeilinger}}]{Resch_PRL05:ThreePhotonStateTomography}%
  \BibitemOpen
  \bibfield  {author} {\bibinfo {author} {\bibfnamefont {K.~J.}\ \bibnamefont
  {Resch}}, \bibinfo {author} {\bibfnamefont {P.}~\bibnamefont {Walther}},\
  and\ \bibinfo {author} {\bibfnamefont {A.}~\bibnamefont {Zeilinger}},\ }\href
  {https://doi.org/10.1103/PhysRevLett.94.070402} {\bibfield  {journal}
  {\bibinfo  {journal} {Phys. Rev. Lett.}\ }\textbf {\bibinfo {volume} {94}},\
  \bibinfo {pages} {070402} (\bibinfo {year} {2005})}\BibitemShut {NoStop}%
\bibitem [{\citenamefont {S\"{o}derholm}\ \emph {et~al.}(2012)\citenamefont
  {S\"{o}derholm}, \citenamefont {Bj\"{o}rk}, \citenamefont {Klimov},
  \citenamefont {S\'{a}nchez-Soto},\ and\ \citenamefont
  {Leuchs}}]{Soderholm_NJP12:PolaTomography}%
  \BibitemOpen
  \bibfield  {author} {\bibinfo {author} {\bibfnamefont {J.}~\bibnamefont
  {S\"{o}derholm}}, \bibinfo {author} {\bibfnamefont {G.}~\bibnamefont
  {Bj\"{o}rk}}, \bibinfo {author} {\bibfnamefont {A.~B.}\ \bibnamefont
  {Klimov}}, \bibinfo {author} {\bibfnamefont {L.~L.}\ \bibnamefont
  {S\'{a}nchez-Soto}},\ and\ \bibinfo {author} {\bibfnamefont {G.}~\bibnamefont
  {Leuchs}},\ }\href {http://stacks.iop.org/1367-2630/14/i=11/a=115014}
  {\bibfield  {journal} {\bibinfo  {journal} {New J. Phys.}\ }\textbf {\bibinfo
  {volume} {14}},\ \bibinfo {pages} {115014} (\bibinfo {year}
  {2012})}\BibitemShut {NoStop}%
\bibitem [{\citenamefont {Sych}\ \emph {et~al.}(2012)\citenamefont {Sych},
  \citenamefont {\ifmmode \check{R}\else \v{R}\fi{}eh\'a\ifmmode~\check{c}\else
  \v{c}\fi{}ek}, \citenamefont {Hradil}, \citenamefont {Leuchs},\ and\
  \citenamefont {S\'anchez-Soto}}]{Sych_PRA12:InfoCompleteContVariable}%
  \BibitemOpen
  \bibfield  {author} {\bibinfo {author} {\bibfnamefont {D.}~\bibnamefont
  {Sych}}, \bibinfo {author} {\bibfnamefont {J.}~\bibnamefont {\ifmmode
  \check{R}\else \v{R}\fi{}eh\'a\ifmmode~\check{c}\else \v{c}\fi{}ek}},
  \bibinfo {author} {\bibfnamefont {Z.}~\bibnamefont {Hradil}}, \bibinfo
  {author} {\bibfnamefont {G.}~\bibnamefont {Leuchs}},\ and\ \bibinfo {author}
  {\bibfnamefont {L.~L.}\ \bibnamefont {S\'anchez-Soto}},\ }\href
  {https://doi.org/10.1103/PhysRevA.86.052123} {\bibfield  {journal} {\bibinfo
  {journal} {Phys. Rev. A}\ }\textbf {\bibinfo {volume} {86}},\ \bibinfo
  {pages} {052123} (\bibinfo {year} {2012})}\BibitemShut {NoStop}%
\bibitem [{\citenamefont {Lvovsky}\ \emph {et~al.}(2001)\citenamefont
  {Lvovsky}, \citenamefont {Hansen}, \citenamefont {Aichele}, \citenamefont
  {Benson}, \citenamefont {Mlynek},\ and\ \citenamefont
  {Schiller}}]{PhysRevLett.87.050402}%
  \BibitemOpen
  \bibfield  {author} {\bibinfo {author} {\bibfnamefont {A.~I.}\ \bibnamefont
  {Lvovsky}}, \bibinfo {author} {\bibfnamefont {H.}~\bibnamefont {Hansen}},
  \bibinfo {author} {\bibfnamefont {T.}~\bibnamefont {Aichele}}, \bibinfo
  {author} {\bibfnamefont {O.}~\bibnamefont {Benson}}, \bibinfo {author}
  {\bibfnamefont {J.}~\bibnamefont {Mlynek}},\ and\ \bibinfo {author}
  {\bibfnamefont {S.}~\bibnamefont {Schiller}},\ }\href
  {https://doi.org/10.1103/PhysRevLett.87.050402} {\bibfield  {journal}
  {\bibinfo  {journal} {Phys. Rev. Lett.}\ }\textbf {\bibinfo {volume} {87}},\
  \bibinfo {pages} {050402} (\bibinfo {year} {2001})}\BibitemShut {NoStop}%
\bibitem [{\citenamefont {Beck}(2000)}]{Beck_PRL00:QuanTomoDetArray}%
  \BibitemOpen
  \bibfield  {author} {\bibinfo {author} {\bibfnamefont {M.}~\bibnamefont
  {Beck}},\ }\href {https://doi.org/10.1103/PhysRevLett.84.5748} {\bibfield
  {journal} {\bibinfo  {journal} {Phys. Rev. Lett.}\ }\textbf {\bibinfo
  {volume} {84}},\ \bibinfo {pages} {5748} (\bibinfo {year}
  {2000})}\BibitemShut {NoStop}%
\bibitem [{\citenamefont {Paris}\ and\ \citenamefont
  {Rehacek}(2004)}]{paris2004quantum}%
  \BibitemOpen
  \bibfield  {author} {\bibinfo {author} {\bibfnamefont {M.}~\bibnamefont
  {Paris}}\ and\ \bibinfo {author} {\bibfnamefont {J.}~\bibnamefont
  {Rehacek}},\ }\href@noop {} {\emph {\bibinfo {title} {Quantum state
  estimation}}}\ (\bibinfo  {publisher} {Springer Science \& Business Media},\
  \bibinfo {year} {2004})\BibitemShut {NoStop}%
\bibitem [{\citenamefont {Beck}\ \emph {et~al.}(2001)\citenamefont {Beck},
  \citenamefont {Dorrer},\ and\ \citenamefont
  {Walmsley}}]{Beck_PRL01:QuanMeasureDetArray}%
  \BibitemOpen
  \bibfield  {author} {\bibinfo {author} {\bibfnamefont {M.}~\bibnamefont
  {Beck}}, \bibinfo {author} {\bibfnamefont {C.}~\bibnamefont {Dorrer}},\ and\
  \bibinfo {author} {\bibfnamefont {I.~A.}\ \bibnamefont {Walmsley}},\ }\href
  {https://doi.org/10.1103/PhysRevLett.87.253601} {\bibfield  {journal}
  {\bibinfo  {journal} {Phys. Rev. Lett.}\ }\textbf {\bibinfo {volume} {87}},\
  \bibinfo {pages} {253601} (\bibinfo {year} {2001})}\BibitemShut {NoStop}%
\bibitem [{\citenamefont {Dawes}\ \emph {et~al.}(2003)\citenamefont {Dawes},
  \citenamefont {Beck},\ and\ \citenamefont
  {Banaszek}}]{Dawes_PRA03:quantumTomowDetArray}%
  \BibitemOpen
  \bibfield  {author} {\bibinfo {author} {\bibfnamefont {A.~M.}\ \bibnamefont
  {Dawes}}, \bibinfo {author} {\bibfnamefont {M.}~\bibnamefont {Beck}},\ and\
  \bibinfo {author} {\bibfnamefont {K.}~\bibnamefont {Banaszek}},\ }\href
  {https://doi.org/10.1103/PhysRevA.67.032102} {\bibfield  {journal} {\bibinfo
  {journal} {Phys. Rev. A}\ }\textbf {\bibinfo {volume} {67}},\ \bibinfo
  {pages} {032102} (\bibinfo {year} {2003})}\BibitemShut {NoStop}%
\bibitem [{\citenamefont {Smith}\ \emph {et~al.}(2005)\citenamefont {Smith},
  \citenamefont {Killett}, \citenamefont {Raymer}, \citenamefont {Walmsley},\
  and\ \citenamefont {Banaszek}}]{Smith_OL05:Measure_SglPhoton}%
  \BibitemOpen
  \bibfield  {author} {\bibinfo {author} {\bibfnamefont {B.~J.}\ \bibnamefont
  {Smith}}, \bibinfo {author} {\bibfnamefont {B.}~\bibnamefont {Killett}},
  \bibinfo {author} {\bibfnamefont {M.~G.}\ \bibnamefont {Raymer}}, \bibinfo
  {author} {\bibfnamefont {I.~A.}\ \bibnamefont {Walmsley}},\ and\ \bibinfo
  {author} {\bibfnamefont {K.}~\bibnamefont {Banaszek}},\ }\href
  {https://doi.org/10.1364/OL.30.003365} {\bibfield  {journal} {\bibinfo
  {journal} {Opt. Lett.}\ }\textbf {\bibinfo {volume} {30}},\ \bibinfo {pages}
  {3365} (\bibinfo {year} {2005})}\BibitemShut {NoStop}%
\bibitem [{\citenamefont {Smithey}\ \emph {et~al.}(1993)\citenamefont
  {Smithey}, \citenamefont {Beck}, \citenamefont {Raymer},\ and\ \citenamefont
  {Faridani}}]{PhysRevLett.70.1244}%
  \BibitemOpen
  \bibfield  {author} {\bibinfo {author} {\bibfnamefont {D.~T.}\ \bibnamefont
  {Smithey}}, \bibinfo {author} {\bibfnamefont {M.}~\bibnamefont {Beck}},
  \bibinfo {author} {\bibfnamefont {M.~G.}\ \bibnamefont {Raymer}},\ and\
  \bibinfo {author} {\bibfnamefont {A.}~\bibnamefont {Faridani}},\ }\href
  {https://doi.org/10.1103/PhysRevLett.70.1244} {\bibfield  {journal} {\bibinfo
   {journal} {Phys. Rev. Lett.}\ }\textbf {\bibinfo {volume} {70}},\ \bibinfo
  {pages} {1244} (\bibinfo {year} {1993})}\BibitemShut {NoStop}%
\bibitem [{\citenamefont {Lvovsky}\ and\ \citenamefont
  {Raymer}(2009)}]{RevModPhys.81.299}%
  \BibitemOpen
  \bibfield  {author} {\bibinfo {author} {\bibfnamefont {A.~I.}\ \bibnamefont
  {Lvovsky}}\ and\ \bibinfo {author} {\bibfnamefont {M.~G.}\ \bibnamefont
  {Raymer}},\ }\href {https://doi.org/10.1103/RevModPhys.81.299} {\bibfield
  {journal} {\bibinfo  {journal} {Rev. Mod. Phys.}\ }\textbf {\bibinfo {volume}
  {81}},\ \bibinfo {pages} {299} (\bibinfo {year} {2009})}\BibitemShut
  {NoStop}%
\bibitem [{\citenamefont {Lundeen}\ \emph {et~al.}(2011)\citenamefont
  {Lundeen}, \citenamefont {Sutherland}, \citenamefont {Patel}, \citenamefont
  {Stewart},\ and\ \citenamefont {Bamber}}]{Lundeen_Nat11:DirectMeasure}%
  \BibitemOpen
  \bibfield  {author} {\bibinfo {author} {\bibfnamefont {J.~S.}\ \bibnamefont
  {Lundeen}}, \bibinfo {author} {\bibfnamefont {B.}~\bibnamefont {Sutherland}},
  \bibinfo {author} {\bibfnamefont {A.}~\bibnamefont {Patel}}, \bibinfo
  {author} {\bibfnamefont {C.}~\bibnamefont {Stewart}},\ and\ \bibinfo {author}
  {\bibfnamefont {C.}~\bibnamefont {Bamber}},\ }\href@noop {} {\bibfield
  {journal} {\bibinfo  {journal} {Nature}\ }\textbf {\bibinfo {volume} {474}},\
  \bibinfo {pages} {188} (\bibinfo {year} {2011})}\BibitemShut {NoStop}%
\bibitem [{\citenamefont {Aharonov}\ \emph {et~al.}(1988)\citenamefont
  {Aharonov}, \citenamefont {Albert},\ and\ \citenamefont
  {Vaidman}}]{Aharonov_PRL88:WeakMeasurementSpin}%
  \BibitemOpen
  \bibfield  {author} {\bibinfo {author} {\bibfnamefont {Y.}~\bibnamefont
  {Aharonov}}, \bibinfo {author} {\bibfnamefont {D.~Z.}\ \bibnamefont
  {Albert}},\ and\ \bibinfo {author} {\bibfnamefont {L.}~\bibnamefont
  {Vaidman}},\ }\href {https://doi.org/10.1103/PhysRevLett.60.1351} {\bibfield
  {journal} {\bibinfo  {journal} {Phys. Rev. Lett.}\ }\textbf {\bibinfo
  {volume} {60}},\ \bibinfo {pages} {1351} (\bibinfo {year}
  {1988})}\BibitemShut {NoStop}%
\bibitem [{\citenamefont {Duck}\ \emph {et~al.}(1989)\citenamefont {Duck},
  \citenamefont {Stevenson},\ and\ \citenamefont
  {Sudarshan}}]{Duck_PRD89:WeakMeasurement}%
  \BibitemOpen
  \bibfield  {author} {\bibinfo {author} {\bibfnamefont {I.~M.}\ \bibnamefont
  {Duck}}, \bibinfo {author} {\bibfnamefont {P.~M.}\ \bibnamefont
  {Stevenson}},\ and\ \bibinfo {author} {\bibfnamefont {E.~C.~G.}\ \bibnamefont
  {Sudarshan}},\ }\href {https://doi.org/10.1103/PhysRevD.40.2112} {\bibfield
  {journal} {\bibinfo  {journal} {Phys. Rev. D}\ }\textbf {\bibinfo {volume}
  {40}},\ \bibinfo {pages} {2112} (\bibinfo {year} {1989})}\BibitemShut
  {NoStop}%
\bibitem [{\citenamefont {Ritchie}\ \emph {et~al.}(1991)\citenamefont
  {Ritchie}, \citenamefont {Story},\ and\ \citenamefont
  {Hulet}}]{Ritchie_PRL91:RealizationWeakMeasurement}%
  \BibitemOpen
  \bibfield  {author} {\bibinfo {author} {\bibfnamefont {N.~W.~M.}\
  \bibnamefont {Ritchie}}, \bibinfo {author} {\bibfnamefont {J.~G.}\
  \bibnamefont {Story}},\ and\ \bibinfo {author} {\bibfnamefont {R.~G.}\
  \bibnamefont {Hulet}},\ }\href {https://doi.org/10.1103/PhysRevLett.66.1107}
  {\bibfield  {journal} {\bibinfo  {journal} {Phys. Rev. Lett.}\ }\textbf
  {\bibinfo {volume} {66}},\ \bibinfo {pages} {1107} (\bibinfo {year}
  {1991})}\BibitemShut {NoStop}%
\bibitem [{\citenamefont
  {Johansen}(2004)}]{Johansen_PRL04:WeakMeasurementArbitraryProbeStates}%
  \BibitemOpen
  \bibfield  {author} {\bibinfo {author} {\bibfnamefont {L.~M.}\ \bibnamefont
  {Johansen}},\ }\href {https://doi.org/10.1103/PhysRevLett.93.120402}
  {\bibfield  {journal} {\bibinfo  {journal} {Phys. Rev. Lett.}\ }\textbf
  {\bibinfo {volume} {93}},\ \bibinfo {pages} {120402} (\bibinfo {year}
  {2004})}\BibitemShut {NoStop}%
\bibitem [{\citenamefont {Hosten}\ and\ \citenamefont
  {Kwiat}(2004)}]{Hosten_Science04:SpinHallEff_weakMeasure}%
  \BibitemOpen
  \bibfield  {author} {\bibinfo {author} {\bibfnamefont {O.}~\bibnamefont
  {Hosten}}\ and\ \bibinfo {author} {\bibfnamefont {P.}~\bibnamefont {Kwiat}},\
  }\href@noop {} {\bibfield  {journal} {\bibinfo  {journal} {Science}\ }\textbf
  {\bibinfo {volume} {319}},\ \bibinfo {pages} {787} (\bibinfo {year}
  {2004})}\BibitemShut {NoStop}%
\bibitem [{\citenamefont {Solli}\ \emph {et~al.}(2004)\citenamefont {Solli},
  \citenamefont {McCormick}, \citenamefont {Chiao}, \citenamefont {Popescu},\
  and\ \citenamefont {Hickmann}}]{Solli_PRL04:FLSL_GeneralizedWeakValue}%
  \BibitemOpen
  \bibfield  {author} {\bibinfo {author} {\bibfnamefont {D.~R.}\ \bibnamefont
  {Solli}}, \bibinfo {author} {\bibfnamefont {C.~F.}\ \bibnamefont
  {McCormick}}, \bibinfo {author} {\bibfnamefont {R.~Y.}\ \bibnamefont
  {Chiao}}, \bibinfo {author} {\bibfnamefont {S.}~\bibnamefont {Popescu}},\
  and\ \bibinfo {author} {\bibfnamefont {J.~M.}\ \bibnamefont {Hickmann}},\
  }\href {https://doi.org/10.1103/PhysRevLett.92.043601} {\bibfield  {journal}
  {\bibinfo  {journal} {Phys. Rev. Lett.}\ }\textbf {\bibinfo {volume} {92}},\
  \bibinfo {pages} {043601} (\bibinfo {year} {2004})}\BibitemShut {NoStop}%
\bibitem [{\citenamefont {Dixon}\ \emph {et~al.}(2009)\citenamefont {Dixon},
  \citenamefont {Starling}, \citenamefont {Jordan},\ and\ \citenamefont
  {Howell}}]{DixonPRL09:WeakDeflection}%
  \BibitemOpen
  \bibfield  {author} {\bibinfo {author} {\bibfnamefont {P.~B.}\ \bibnamefont
  {Dixon}}, \bibinfo {author} {\bibfnamefont {D.~J.}\ \bibnamefont {Starling}},
  \bibinfo {author} {\bibfnamefont {A.~N.}\ \bibnamefont {Jordan}},\ and\
  \bibinfo {author} {\bibfnamefont {J.~C.}\ \bibnamefont {Howell}},\ }\href
  {https://doi.org/10.1103/PhysRevLett.102.173601} {\bibfield  {journal}
  {\bibinfo  {journal} {Phys. Rev. Lett.}\ }\textbf {\bibinfo {volume} {102}},\
  \bibinfo {pages} {173601} (\bibinfo {year} {2009})}\BibitemShut {NoStop}%
\bibitem [{\citenamefont {Feizpour}\ \emph {et~al.}(2011)\citenamefont
  {Feizpour}, \citenamefont {Xing},\ and\ \citenamefont
  {Steinberg}}]{Feizpour_PRL11:WeakSinglePhotonNonlinearity}%
  \BibitemOpen
  \bibfield  {author} {\bibinfo {author} {\bibfnamefont {A.}~\bibnamefont
  {Feizpour}}, \bibinfo {author} {\bibfnamefont {X.}~\bibnamefont {Xing}},\
  and\ \bibinfo {author} {\bibfnamefont {A.~M.}\ \bibnamefont {Steinberg}},\
  }\href {https://doi.org/10.1103/PhysRevLett.107.133603} {\bibfield  {journal}
  {\bibinfo  {journal} {Phys. Rev. Lett.}\ }\textbf {\bibinfo {volume} {107}},\
  \bibinfo {pages} {133603} (\bibinfo {year} {2011})}\BibitemShut {NoStop}%
\bibitem [{\citenamefont {Kocsis}\ \emph {et~al.}(2011)\citenamefont {Kocsis},
  \citenamefont {Braverman}, \citenamefont {Ravets}, \citenamefont {Stevens},
  \citenamefont {Mirin}, \citenamefont {Shalm},\ and\ \citenamefont
  {Steinberg}}]{Kocsis_Science12:EMCCD_entanglement}%
  \BibitemOpen
  \bibfield  {author} {\bibinfo {author} {\bibfnamefont {S.}~\bibnamefont
  {Kocsis}}, \bibinfo {author} {\bibfnamefont {B.}~\bibnamefont {Braverman}},
  \bibinfo {author} {\bibfnamefont {S.}~\bibnamefont {Ravets}}, \bibinfo
  {author} {\bibfnamefont {M.~J.}\ \bibnamefont {Stevens}}, \bibinfo {author}
  {\bibfnamefont {R.~P.}\ \bibnamefont {Mirin}}, \bibinfo {author}
  {\bibfnamefont {L.~K.}\ \bibnamefont {Shalm}},\ and\ \bibinfo {author}
  {\bibfnamefont {A.~M.}\ \bibnamefont {Steinberg}},\ }\href
  {https://doi.org/10.1126/science.1202218} {\bibfield  {journal} {\bibinfo
  {journal} {Science}\ }\textbf {\bibinfo {volume} {332}},\ \bibinfo {pages}
  {1170} (\bibinfo {year} {2011})}\BibitemShut {NoStop}%
\bibitem [{\citenamefont {Dressel}\ \emph {et~al.}(2014)\citenamefont
  {Dressel}, \citenamefont {Malik}, \citenamefont {Miatto}, \citenamefont
  {Jordan},\ and\ \citenamefont {Boyd}}]{Dressel_13ti:WeakReview}%
  \BibitemOpen
  \bibfield  {author} {\bibinfo {author} {\bibfnamefont {J.}~\bibnamefont
  {Dressel}}, \bibinfo {author} {\bibfnamefont {M.}~\bibnamefont {Malik}},
  \bibinfo {author} {\bibfnamefont {F.~M.}\ \bibnamefont {Miatto}}, \bibinfo
  {author} {\bibfnamefont {A.~N.}\ \bibnamefont {Jordan}},\ and\ \bibinfo
  {author} {\bibfnamefont {R.~W.}\ \bibnamefont {Boyd}},\ }\href@noop {}
  {\bibfield  {journal} {\bibinfo  {journal} {Rev. Mod. Phys.}\ }\textbf
  {\bibinfo {volume} {86}},\ \bibinfo {pages} {307} (\bibinfo {year}
  {2014})}\BibitemShut {NoStop}%
\bibitem [{\citenamefont {Bamber}\ and\ \citenamefont
  {Lundeen}(2014)}]{PhysRevLett.112.070405}%
  \BibitemOpen
  \bibfield  {author} {\bibinfo {author} {\bibfnamefont {C.}~\bibnamefont
  {Bamber}}\ and\ \bibinfo {author} {\bibfnamefont {J.~S.}\ \bibnamefont
  {Lundeen}},\ }\href {https://doi.org/10.1103/PhysRevLett.112.070405}
  {\bibfield  {journal} {\bibinfo  {journal} {Phys. Rev. Lett.}\ }\textbf
  {\bibinfo {volume} {112}},\ \bibinfo {pages} {070405} (\bibinfo {year}
  {2014})}\BibitemShut {NoStop}%
\bibitem [{\citenamefont {Turek}\ \emph {et~al.}(2015)\citenamefont {Turek},
  \citenamefont {Kobayashi}, \citenamefont {Akutsu}, \citenamefont {Sun},\ and\
  \citenamefont {Shikano}}]{turek2015post}%
  \BibitemOpen
  \bibfield  {author} {\bibinfo {author} {\bibfnamefont {Y.}~\bibnamefont
  {Turek}}, \bibinfo {author} {\bibfnamefont {H.}~\bibnamefont {Kobayashi}},
  \bibinfo {author} {\bibfnamefont {T.}~\bibnamefont {Akutsu}}, \bibinfo
  {author} {\bibfnamefont {C.}~\bibnamefont {Sun}},\ and\ \bibinfo {author}
  {\bibfnamefont {Y.}~\bibnamefont {Shikano}},\ }\href@noop {} {\bibfield
  {journal} {\bibinfo  {journal} {New J. Phys.}\ }\textbf {\bibinfo {volume}
  {17}},\ \bibinfo {pages} {083029} (\bibinfo {year} {2015})}\BibitemShut
  {NoStop}%
\bibitem [{\citenamefont {Ren}\ \emph {et~al.}(2019)\citenamefont {Ren},
  \citenamefont {Wang},\ and\ \citenamefont {Du}}]{ren2019efficient}%
  \BibitemOpen
  \bibfield  {author} {\bibinfo {author} {\bibfnamefont {C.}~\bibnamefont
  {Ren}}, \bibinfo {author} {\bibfnamefont {Y.}~\bibnamefont {Wang}},\ and\
  \bibinfo {author} {\bibfnamefont {J.}~\bibnamefont {Du}},\ }\href@noop {}
  {\bibfield  {journal} {\bibinfo  {journal} {Phys. Rev. Applied}\ }\textbf
  {\bibinfo {volume} {12}},\ \bibinfo {pages} {014045} (\bibinfo {year}
  {2019})}\BibitemShut {NoStop}%
\bibitem [{\citenamefont {Ogawa}\ \emph {et~al.}(2019)\citenamefont {Ogawa},
  \citenamefont {Yasuhiko}, \citenamefont {Kobayashi}, \citenamefont
  {Nakanishi},\ and\ \citenamefont {Tomita}}]{ogawa2019framework}%
  \BibitemOpen
  \bibfield  {author} {\bibinfo {author} {\bibfnamefont {K.}~\bibnamefont
  {Ogawa}}, \bibinfo {author} {\bibfnamefont {O.}~\bibnamefont {Yasuhiko}},
  \bibinfo {author} {\bibfnamefont {H.}~\bibnamefont {Kobayashi}}, \bibinfo
  {author} {\bibfnamefont {T.}~\bibnamefont {Nakanishi}},\ and\ \bibinfo
  {author} {\bibfnamefont {A.}~\bibnamefont {Tomita}},\ }\href@noop {}
  {\bibfield  {journal} {\bibinfo  {journal} {New J. Phys.}\ }\textbf {\bibinfo
  {volume} {21}},\ \bibinfo {pages} {043013} (\bibinfo {year}
  {2019})}\BibitemShut {NoStop}%
\bibitem [{\citenamefont {Vallone}\ and\ \citenamefont
  {Dequal}(2016)}]{vallone2016strong}%
  \BibitemOpen
  \bibfield  {author} {\bibinfo {author} {\bibfnamefont {G.}~\bibnamefont
  {Vallone}}\ and\ \bibinfo {author} {\bibfnamefont {D.}~\bibnamefont
  {Dequal}},\ }\href@noop {} {\bibfield  {journal} {\bibinfo  {journal} {Phys.
  Rev. Lett.}\ }\textbf {\bibinfo {volume} {116}},\ \bibinfo {pages} {040502}
  (\bibinfo {year} {2016})}\BibitemShut {NoStop}%
\bibitem [{\citenamefont {Zou}\ \emph {et~al.}(2015)\citenamefont {Zou},
  \citenamefont {Zhang},\ and\ \citenamefont {Song}}]{zou2015direct}%
  \BibitemOpen
  \bibfield  {author} {\bibinfo {author} {\bibfnamefont {P.}~\bibnamefont
  {Zou}}, \bibinfo {author} {\bibfnamefont {Z.-M.}\ \bibnamefont {Zhang}},\
  and\ \bibinfo {author} {\bibfnamefont {W.}~\bibnamefont {Song}},\ }\href@noop
  {} {\bibfield  {journal} {\bibinfo  {journal} {Phys. Rev. A}\ }\textbf
  {\bibinfo {volume} {91}},\ \bibinfo {pages} {052109} (\bibinfo {year}
  {2015})}\BibitemShut {NoStop}%
\bibitem [{\citenamefont {Zhang}\ \emph {et~al.}(2020)\citenamefont {Zhang},
  \citenamefont {Hu}, \citenamefont {Hou}, \citenamefont {Tang}, \citenamefont
  {Zhu}, \citenamefont {Xiang}, \citenamefont {Li}, \citenamefont {Guo},\ and\
  \citenamefont {Zhang}}]{PhysRevA.101.012119}%
  \BibitemOpen
  \bibfield  {author} {\bibinfo {author} {\bibfnamefont {C.-R.}\ \bibnamefont
  {Zhang}}, \bibinfo {author} {\bibfnamefont {M.-J.}\ \bibnamefont {Hu}},
  \bibinfo {author} {\bibfnamefont {Z.-B.}\ \bibnamefont {Hou}}, \bibinfo
  {author} {\bibfnamefont {J.-F.}\ \bibnamefont {Tang}}, \bibinfo {author}
  {\bibfnamefont {J.}~\bibnamefont {Zhu}}, \bibinfo {author} {\bibfnamefont
  {G.-Y.}\ \bibnamefont {Xiang}}, \bibinfo {author} {\bibfnamefont {C.-F.}\
  \bibnamefont {Li}}, \bibinfo {author} {\bibfnamefont {G.-C.}\ \bibnamefont
  {Guo}},\ and\ \bibinfo {author} {\bibfnamefont {Y.-S.}\ \bibnamefont
  {Zhang}},\ }\href {https://doi.org/10.1103/PhysRevA.101.012119} {\bibfield
  {journal} {\bibinfo  {journal} {Phys. Rev. A}\ }\textbf {\bibinfo {volume}
  {101}},\ \bibinfo {pages} {012119} (\bibinfo {year} {2020})}\BibitemShut
  {NoStop}%
\bibitem [{\citenamefont {Thekkadath}\ \emph {et~al.}(2016)\citenamefont
  {Thekkadath}, \citenamefont {Giner}, \citenamefont {Chalich}, \citenamefont
  {Horton}, \citenamefont {Banker},\ and\ \citenamefont
  {Lundeen}}]{Thekkadath2016Direct}%
  \BibitemOpen
  \bibfield  {author} {\bibinfo {author} {\bibfnamefont {G.~S.}\ \bibnamefont
  {Thekkadath}}, \bibinfo {author} {\bibfnamefont {L.}~\bibnamefont {Giner}},
  \bibinfo {author} {\bibfnamefont {Y.}~\bibnamefont {Chalich}}, \bibinfo
  {author} {\bibfnamefont {M.~J.}\ \bibnamefont {Horton}}, \bibinfo {author}
  {\bibfnamefont {J.}~\bibnamefont {Banker}},\ and\ \bibinfo {author}
  {\bibfnamefont {J.~S.}\ \bibnamefont {Lundeen}},\ }\href
  {https://doi.org/10.1103/PhysRevLett.117.120401} {\bibfield  {journal}
  {\bibinfo  {journal} {Phys. Rev. Lett.}\ }\textbf {\bibinfo {volume} {117}},\
  \bibinfo {pages} {120401} (\bibinfo {year} {2016})}\BibitemShut {NoStop}%
\bibitem [{\citenamefont {Lundeen}\ and\ \citenamefont
  {Bamber}(2012)}]{Lundeen_PRL12:WeakMeasureGeneral}%
  \BibitemOpen
  \bibfield  {author} {\bibinfo {author} {\bibfnamefont {J.~S.}\ \bibnamefont
  {Lundeen}}\ and\ \bibinfo {author} {\bibfnamefont {C.}~\bibnamefont
  {Bamber}},\ }\href {https://doi.org/10.1103/PhysRevLett.108.070402}
  {\bibfield  {journal} {\bibinfo  {journal} {Phys. Rev. Lett.}\ }\textbf
  {\bibinfo {volume} {108}},\ \bibinfo {pages} {070402} (\bibinfo {year}
  {2012})}\BibitemShut {NoStop}%
\bibitem [{\citenamefont {Shengjun}(2013)}]{Wu_SciRep:StateTomography}%
  \BibitemOpen
  \bibfield  {author} {\bibinfo {author} {\bibfnamefont {W.}~\bibnamefont
  {Shengjun}},\ }\href {https://doi.org/10.1038/srep01193} {\bibfield
  {journal} {\bibinfo  {journal} {Sci. Rep.}\ }\textbf {\bibinfo {volume}
  {3}},\ \bibinfo {pages} {1193} (\bibinfo {year} {2013})}\BibitemShut
  {NoStop}%
\bibitem [{\citenamefont {Salvail}\ \emph {et~al.}(2013)\citenamefont
  {Salvail}, \citenamefont {Agnew}, \citenamefont {Johnson}, \citenamefont
  {Bolduc}, \citenamefont {Leach},\ and\ \citenamefont
  {Boyd}}]{Saivail_NatPhon13:PolarizationDirectMeasure}%
  \BibitemOpen
  \bibfield  {author} {\bibinfo {author} {\bibfnamefont {J.~Z.}\ \bibnamefont
  {Salvail}}, \bibinfo {author} {\bibfnamefont {M.}~\bibnamefont {Agnew}},
  \bibinfo {author} {\bibfnamefont {A.~S.}\ \bibnamefont {Johnson}}, \bibinfo
  {author} {\bibfnamefont {E.}~\bibnamefont {Bolduc}}, \bibinfo {author}
  {\bibfnamefont {J.}~\bibnamefont {Leach}},\ and\ \bibinfo {author}
  {\bibfnamefont {R.~W.}\ \bibnamefont {Boyd}},\ }\href@noop {} {\bibfield
  {journal} {\bibinfo  {journal} {Nat. Photon.}\ }\textbf {\bibinfo {volume}
  {7}},\ \bibinfo {pages} {316} (\bibinfo {year} {2013})}\BibitemShut {NoStop}%
\bibitem [{\citenamefont {Mirhosseini}\ \emph {et~al.}(2014)\citenamefont
  {Mirhosseini}, \citenamefont {Maga\~na{-}Loaiza}, \citenamefont
  {Hashemi~Rafsanjani},\ and\ \citenamefont
  {Boyd}}]{Mirhosseini_PRL14:CompressiveDM}%
  \BibitemOpen
  \bibfield  {author} {\bibinfo {author} {\bibfnamefont {M.}~\bibnamefont
  {Mirhosseini}}, \bibinfo {author} {\bibfnamefont {O.~S.}\ \bibnamefont
  {Maga\~na{-}Loaiza}}, \bibinfo {author} {\bibfnamefont {S.~M.}\ \bibnamefont
  {Hashemi~Rafsanjani}},\ and\ \bibinfo {author} {\bibfnamefont {R.~W.}\
  \bibnamefont {Boyd}},\ }\href
  {https://doi.org/10.1103/PhysRevLett.113.090402} {\bibfield  {journal}
  {\bibinfo  {journal} {Phys. Rev. Lett.}\ }\textbf {\bibinfo {volume} {113}},\
  \bibinfo {pages} {090402} (\bibinfo {year} {2014})}\BibitemShut {NoStop}%
\bibitem [{\citenamefont {Malik}\ \emph {et~al.}(2014)\citenamefont {Malik},
  \citenamefont {Mirhosseini}, \citenamefont {Lavery}, \citenamefont {Leach},
  \citenamefont {Padgett},\ and\ \citenamefont
  {Boyd}}]{Malik_13tj:DirectMeasureOAM}%
  \BibitemOpen
  \bibfield  {author} {\bibinfo {author} {\bibfnamefont {M.}~\bibnamefont
  {Malik}}, \bibinfo {author} {\bibfnamefont {M.}~\bibnamefont {Mirhosseini}},
  \bibinfo {author} {\bibfnamefont {M.~P.~J.}\ \bibnamefont {Lavery}}, \bibinfo
  {author} {\bibfnamefont {J.}~\bibnamefont {Leach}}, \bibinfo {author}
  {\bibfnamefont {M.~J.}\ \bibnamefont {Padgett}},\ and\ \bibinfo {author}
  {\bibfnamefont {R.~W.}\ \bibnamefont {Boyd}},\ }\href
  {https://doi.org/10.1038/ncomms4115} {\bibfield  {journal} {\bibinfo
  {journal} {Nat. Commun.}\ }\textbf {\bibinfo {volume} {4}},\ \bibinfo {pages}
  {3115} (\bibinfo {year} {2014})}\BibitemShut {NoStop}%
\bibitem [{\citenamefont {Mirhosseini}\ \emph {et~al.}(2016)\citenamefont
  {Mirhosseini}, \citenamefont {Maga\~na{-}Loaiza}, \citenamefont {Chen},
  \citenamefont {Hashemi~Rafsanjani},\ and\ \citenamefont
  {Boyd}}]{Mirhosseini2016Wigner}%
  \BibitemOpen
  \bibfield  {author} {\bibinfo {author} {\bibfnamefont {M.}~\bibnamefont
  {Mirhosseini}}, \bibinfo {author} {\bibfnamefont {O.~S.}\ \bibnamefont
  {Maga\~na{-}Loaiza}}, \bibinfo {author} {\bibfnamefont {C.}~\bibnamefont
  {Chen}}, \bibinfo {author} {\bibfnamefont {S.~M.}\ \bibnamefont
  {Hashemi~Rafsanjani}},\ and\ \bibinfo {author} {\bibfnamefont {R.~W.}\
  \bibnamefont {Boyd}},\ }\href
  {https://doi.org/10.1103/PhysRevLett.116.130402} {\bibfield  {journal}
  {\bibinfo  {journal} {Phys. Rev. Lett.}\ }\textbf {\bibinfo {volume} {116}},\
  \bibinfo {pages} {130402} (\bibinfo {year} {2016})}\BibitemShut {NoStop}%
\bibitem [{\citenamefont {Liu}\ \emph {et~al.}(2019)\citenamefont {Liu},
  \citenamefont {Long}, \citenamefont {Zhang}, \citenamefont {Lake},
  \citenamefont {Gao}, \citenamefont {Pappas},\ and\ \citenamefont
  {Li}}]{liu2019efficient}%
  \BibitemOpen
  \bibfield  {author} {\bibinfo {author} {\bibfnamefont {R.}~\bibnamefont
  {Liu}}, \bibinfo {author} {\bibfnamefont {J.}~\bibnamefont {Long}}, \bibinfo
  {author} {\bibfnamefont {P.}~\bibnamefont {Zhang}}, \bibinfo {author}
  {\bibfnamefont {R.~E.}\ \bibnamefont {Lake}}, \bibinfo {author}
  {\bibfnamefont {H.}~\bibnamefont {Gao}}, \bibinfo {author} {\bibfnamefont
  {D.~P.}\ \bibnamefont {Pappas}},\ and\ \bibinfo {author} {\bibfnamefont
  {F.}~\bibnamefont {Li}},\ }\href@noop {} {\bibfield  {journal} {\bibinfo
  {journal} {arXiv:1908.00577}\ } (\bibinfo {year} {2019})}\BibitemShut
  {NoStop}%
\bibitem [{\citenamefont {Mandel}\ and\ \citenamefont
  {Wolf}(1995)}]{mandel1995optical}%
  \BibitemOpen
  \bibfield  {author} {\bibinfo {author} {\bibfnamefont {L.}~\bibnamefont
  {Mandel}}\ and\ \bibinfo {author} {\bibfnamefont {E.}~\bibnamefont {Wolf}},\
  }\href@noop {} {\emph {\bibinfo {title} {Optical coherence and quantum
  optics}}}\ (\bibinfo  {publisher} {Cambridge University Press, Cambridge},\
  \bibinfo {year} {1995})\BibitemShut {NoStop}%
\bibitem [{\citenamefont {Hradil}\ \emph {et~al.}(2010)\citenamefont {Hradil},
  \citenamefont {\ifmmode \check{R}\else \v{R}\fi{}eh\'a\ifmmode~\check{c}\else
  \v{c}\fi{}ek},\ and\ \citenamefont {S\'anchez-Soto}}]{hradil2010quantum}%
  \BibitemOpen
  \bibfield  {author} {\bibinfo {author} {\bibfnamefont {Z.}~\bibnamefont
  {Hradil}}, \bibinfo {author} {\bibfnamefont {J.}~\bibnamefont {\ifmmode
  \check{R}\else \v{R}\fi{}eh\'a\ifmmode~\check{c}\else \v{c}\fi{}ek}},\ and\
  \bibinfo {author} {\bibfnamefont {L.~L.}\ \bibnamefont {S\'anchez-Soto}},\
  }\href {https://doi.org/10.1103/PhysRevLett.105.010401} {\bibfield  {journal}
  {\bibinfo  {journal} {Phys. Rev. Lett.}\ }\textbf {\bibinfo {volume} {105}},\
  \bibinfo {pages} {010401} (\bibinfo {year} {2010})}\BibitemShut {NoStop}%
\bibitem [{\citenamefont {Shi}\ \emph {et~al.}(2015)\citenamefont {Shi},
  \citenamefont {Mirhosseini}, \citenamefont {Margiewicz}, \citenamefont
  {Malik}, \citenamefont {Rivera}, \citenamefont {Zhu},\ and\ \citenamefont
  {Boyd}}]{Zhimin2015Scan}%
  \BibitemOpen
  \bibfield  {author} {\bibinfo {author} {\bibfnamefont {Z.}~\bibnamefont
  {Shi}}, \bibinfo {author} {\bibfnamefont {M.}~\bibnamefont {Mirhosseini}},
  \bibinfo {author} {\bibfnamefont {J.}~\bibnamefont {Margiewicz}}, \bibinfo
  {author} {\bibfnamefont {M.}~\bibnamefont {Malik}}, \bibinfo {author}
  {\bibfnamefont {F.}~\bibnamefont {Rivera}}, \bibinfo {author} {\bibfnamefont
  {Z.}~\bibnamefont {Zhu}},\ and\ \bibinfo {author} {\bibfnamefont {R.~W.}\
  \bibnamefont {Boyd}},\ }\href {https://doi.org/10.1364/OPTICA.2.000388}
  {\bibfield  {journal} {\bibinfo  {journal} {Optica}\ }\textbf {\bibinfo
  {volume} {2}},\ \bibinfo {pages} {388} (\bibinfo {year} {2015})}\BibitemShut
  {NoStop}%
\bibitem [{\citenamefont {Zhu}\ \emph {et~al.}(2019)\citenamefont {Zhu},
  \citenamefont {Hay}, \citenamefont {Zhou}, \citenamefont {Fyffe},
  \citenamefont {Kantor}, \citenamefont {Agarwal}, \citenamefont {Boyd},\ and\
  \citenamefont {Shi}}]{Single2019Zhu}%
  \BibitemOpen
  \bibfield  {author} {\bibinfo {author} {\bibfnamefont {Z.}~\bibnamefont
  {Zhu}}, \bibinfo {author} {\bibfnamefont {D.}~\bibnamefont {Hay}}, \bibinfo
  {author} {\bibfnamefont {Y.}~\bibnamefont {Zhou}}, \bibinfo {author}
  {\bibfnamefont {A.}~\bibnamefont {Fyffe}}, \bibinfo {author} {\bibfnamefont
  {B.}~\bibnamefont {Kantor}}, \bibinfo {author} {\bibfnamefont {G.~S.}\
  \bibnamefont {Agarwal}}, \bibinfo {author} {\bibfnamefont {R.~W.}\
  \bibnamefont {Boyd}},\ and\ \bibinfo {author} {\bibfnamefont
  {Z.}~\bibnamefont {Shi}},\ }\href
  {https://doi.org/10.1103/PhysRevApplied.12.034036} {\bibfield  {journal}
  {\bibinfo  {journal} {Phys. Rev. Applied}\ }\textbf {\bibinfo {volume}
  {12}},\ \bibinfo {pages} {034036} (\bibinfo {year} {2019})}\BibitemShut
  {NoStop}%
\bibitem [{\citenamefont {Nielsen}\ and\ \citenamefont
  {Chuang}(2010)}]{nielsenchuang2010}%
  \BibitemOpen
  \bibfield  {author} {\bibinfo {author} {\bibfnamefont {M.~A.}\ \bibnamefont
  {Nielsen}}\ and\ \bibinfo {author} {\bibfnamefont {I.~L.}\ \bibnamefont
  {Chuang}},\ }\href@noop {} {\emph {\bibinfo {title} {Quantum Computation and
  Quantum Information}}}\ (\bibinfo  {publisher} {Cambridge University Press,
  Cambridge},\ \bibinfo {year} {2010})\BibitemShut {NoStop}%
\bibitem [{\citenamefont {Trefethen}\ and\ \citenamefont
  {Bau~III}(1997)}]{trefethen1997numerical}%
  \BibitemOpen
  \bibfield  {author} {\bibinfo {author} {\bibfnamefont {L.~N.}\ \bibnamefont
  {Trefethen}}\ and\ \bibinfo {author} {\bibfnamefont {D.}~\bibnamefont
  {Bau~III}},\ }\href@noop {} {\emph {\bibinfo {title} {Numerical linear
  algebra}}}\ (\bibinfo  {publisher} {SIAM, Philadelphia},\ \bibinfo {year}
  {1997})\BibitemShut {NoStop}%
\bibitem [{\citenamefont {{Klema}}\ and\ \citenamefont
  {{Laub}}(1980)}]{Klema1980singular}%
  \BibitemOpen
  \bibfield  {author} {\bibinfo {author} {\bibfnamefont {V.}~\bibnamefont
  {{Klema}}}\ and\ \bibinfo {author} {\bibfnamefont {A.}~\bibnamefont
  {{Laub}}},\ }\href {https://doi.org/10.1109/TAC.1980.1102314} {\bibfield
  {journal} {\bibinfo  {journal} {IEEE Trans. Automat. Contr.}\ }\textbf
  {\bibinfo {volume} {25}},\ \bibinfo {pages} {164} (\bibinfo {year}
  {1980})}\BibitemShut {NoStop}%
\bibitem [{\citenamefont {Wolf}(1982)}]{Wolf82}%
  \BibitemOpen
  \bibfield  {author} {\bibinfo {author} {\bibfnamefont {E.}~\bibnamefont
  {Wolf}},\ }\href {https://doi.org/10.1364/JOSA.72.000343} {\bibfield
  {journal} {\bibinfo  {journal} {J. Opt. Soc. Am.}\ }\textbf {\bibinfo
  {volume} {72}},\ \bibinfo {pages} {343} (\bibinfo {year} {1982})}\BibitemShut
  {NoStop}%
\bibitem [{\citenamefont {Wolf}(1981)}]{WOLF19813}%
  \BibitemOpen
  \bibfield  {author} {\bibinfo {author} {\bibfnamefont {E.}~\bibnamefont
  {Wolf}},\ }\href
  {https://doi.org/https://doi.org/10.1016/0030-4018(81)90295-9} {\bibfield
  {journal} {\bibinfo  {journal} {Opt. Commun.}\ }\textbf {\bibinfo {volume}
  {38}},\ \bibinfo {pages} {3 } (\bibinfo {year} {1981})}\BibitemShut {NoStop}%
\bibitem [{\citenamefont {Arriz\'{o}n}\ \emph {et~al.}(2007)\citenamefont
  {Arriz\'{o}n}, \citenamefont {Ruiz}, \citenamefont {Carrada},\ and\
  \citenamefont {Gonz\'{a}lez}}]{Arrizon07}%
  \BibitemOpen
  \bibfield  {author} {\bibinfo {author} {\bibfnamefont {V.}~\bibnamefont
  {Arriz\'{o}n}}, \bibinfo {author} {\bibfnamefont {U.}~\bibnamefont {Ruiz}},
  \bibinfo {author} {\bibfnamefont {R.}~\bibnamefont {Carrada}},\ and\ \bibinfo
  {author} {\bibfnamefont {L.~A.}\ \bibnamefont {Gonz\'{a}lez}},\ }\href
  {https://doi.org/10.1364/JOSAA.24.003500} {\bibfield  {journal} {\bibinfo
  {journal} {J. Opt. Soc. Am. A}\ }\textbf {\bibinfo {volume} {24}},\ \bibinfo
  {pages} {3500} (\bibinfo {year} {2007})}\BibitemShut {NoStop}%
\bibitem [{\citenamefont {Rodenburg}\ \emph {et~al.}(2014)\citenamefont
  {Rodenburg}, \citenamefont {Mirhosseini}, \citenamefont {Maga\~na Loaiza},\
  and\ \citenamefont {Boyd}}]{Rodenburg2014Experimental}%
  \BibitemOpen
  \bibfield  {author} {\bibinfo {author} {\bibfnamefont {B.}~\bibnamefont
  {Rodenburg}}, \bibinfo {author} {\bibfnamefont {M.}~\bibnamefont
  {Mirhosseini}}, \bibinfo {author} {\bibfnamefont {O.~S.}\ \bibnamefont
  {Maga\~na Loaiza}},\ and\ \bibinfo {author} {\bibfnamefont {R.~W.}\
  \bibnamefont {Boyd}},\ }\href {https://doi.org/10.1364/JOSAB.31.000A51}
  {\bibfield  {journal} {\bibinfo  {journal} {J. Opt. Soc. Am. B}\ }\textbf
  {\bibinfo {volume} {31}},\ \bibinfo {pages} {A51} (\bibinfo {year}
  {2014})}\BibitemShut {NoStop}%
\bibitem [{\citenamefont {Svelto}\ and\ \citenamefont
  {Hanna}(2010)}]{svelto2010principles}%
  \BibitemOpen
  \bibfield  {author} {\bibinfo {author} {\bibfnamefont {O.}~\bibnamefont
  {Svelto}}\ and\ \bibinfo {author} {\bibfnamefont {D.~C.}\ \bibnamefont
  {Hanna}},\ }\href@noop {} {\emph {\bibinfo {title} {Principles of lasers}}}\
  (\bibinfo  {publisher} {Springer, New York},\ \bibinfo {year}
  {2010})\BibitemShut {NoStop}%
\bibitem [{Gau()}]{GaussFilter}%
  \BibitemOpen
  \href@noop {} {}\bibinfo {note} {In our algorithm, we use a digital Gaussian
  low-pass filtering kernel that has a standard deviation width of 15 pixels.
  Therefore, the effective dimension of the measured photon state is reduced by
  approximately a factor of 15. However, it is worth noting that the use of a
  Gaussian filter is not necessary, and the undesirable fringes caused by dusts
  and glass thin film interference can be minimized alternatively by
  experimental efforts}\BibitemShut {NoStop}%
\bibitem [{\citenamefont {James}\ \emph {et~al.}(2001)\citenamefont {James},
  \citenamefont {Kwiat}, \citenamefont {Munro},\ and\ \citenamefont
  {White}}]{Measurement2001James}%
  \BibitemOpen
  \bibfield  {author} {\bibinfo {author} {\bibfnamefont {D.~F.~V.}\
  \bibnamefont {James}}, \bibinfo {author} {\bibfnamefont {P.~G.}\ \bibnamefont
  {Kwiat}}, \bibinfo {author} {\bibfnamefont {W.~J.}\ \bibnamefont {Munro}},\
  and\ \bibinfo {author} {\bibfnamefont {A.~G.}\ \bibnamefont {White}},\ }\href
  {https://doi.org/10.1103/PhysRevA.64.052312} {\bibfield  {journal} {\bibinfo
  {journal} {Phys. Rev. A}\ }\textbf {\bibinfo {volume} {64}},\ \bibinfo
  {pages} {052312} (\bibinfo {year} {2001})}\BibitemShut {NoStop}%
\bibitem [{\citenamefont {Torlai}\ \emph {et~al.}(2018)\citenamefont {Torlai},
  \citenamefont {Mazzola}, \citenamefont {Carrasquilla}, \citenamefont
  {Troyer}, \citenamefont {Melko},\ and\ \citenamefont
  {Carleo}}]{torlai2018neural}%
  \BibitemOpen
  \bibfield  {author} {\bibinfo {author} {\bibfnamefont {G.}~\bibnamefont
  {Torlai}}, \bibinfo {author} {\bibfnamefont {G.}~\bibnamefont {Mazzola}},
  \bibinfo {author} {\bibfnamefont {J.}~\bibnamefont {Carrasquilla}}, \bibinfo
  {author} {\bibfnamefont {M.}~\bibnamefont {Troyer}}, \bibinfo {author}
  {\bibfnamefont {R.}~\bibnamefont {Melko}},\ and\ \bibinfo {author}
  {\bibfnamefont {G.}~\bibnamefont {Carleo}},\ }\href@noop {} {\bibfield
  {journal} {\bibinfo  {journal} {Nat. Phys.}\ }\textbf {\bibinfo {volume}
  {14}},\ \bibinfo {pages} {447} (\bibinfo {year} {2018})}\BibitemShut
  {NoStop}%
\bibitem [{\citenamefont {Berkhout}\ \emph {et~al.}(2010)\citenamefont
  {Berkhout}, \citenamefont {Lavery}, \citenamefont {Courtial}, \citenamefont
  {Beijersbergen},\ and\ \citenamefont {Padgett}}]{berkhout2010efficient}%
  \BibitemOpen
  \bibfield  {author} {\bibinfo {author} {\bibfnamefont {G.~C.~G.}\
  \bibnamefont {Berkhout}}, \bibinfo {author} {\bibfnamefont {M.~P.~J.}\
  \bibnamefont {Lavery}}, \bibinfo {author} {\bibfnamefont {J.}~\bibnamefont
  {Courtial}}, \bibinfo {author} {\bibfnamefont {M.~W.}\ \bibnamefont
  {Beijersbergen}},\ and\ \bibinfo {author} {\bibfnamefont {M.~J.}\
  \bibnamefont {Padgett}},\ }\href
  {https://doi.org/10.1103/PhysRevLett.105.153601} {\bibfield  {journal}
  {\bibinfo  {journal} {Phys. Rev. Lett.}\ }\textbf {\bibinfo {volume} {105}},\
  \bibinfo {pages} {153601} (\bibinfo {year} {2010})}\BibitemShut {NoStop}%
\bibitem [{Sup()}]{Supplement}%
  \BibitemOpen
  \href@noop {} {}\bibinfo {note} {See Supplemental Material at [URL will be
  inserted by publisher] for a possible realization of the 2D-to-1D beam
  reshaping and the discussion of photon efficiency, which includes Refs.
  \cite{park2017large, Zhimin2015Scan, mandel1995optical,
  Charbon_PTRSLA14:SinglePhotonImagingSPADReview,
  Gariepy_NatComm15:SinglePhotonFlightImaging,
  Dawes_PRA03:quantumTomowDetArray, Kocsis_Science12:EMCCD_entanglement,
  Edgar_NatComm12:EMCCD_entanglement, Lemos_Nat14:QIm_w_undetectedPhotons,
  Tasca_OpEx13:GI_ICCD, Aspden_NJP13:EPRGI_using_ICCD,
  Fickler_SciRep:RealTimeImagingEntanglement,
  Machulka_OpEx14:SpatialPropofTwinbeamCorrelation,
  Morris_NatComm15:ImagingSinglePhotons,
  Chrapkiewicz_NatPhoton16:ImagingSinglePhotons}.}\BibitemShut {Stop}%
\bibitem [{\citenamefont {Park}\ \emph {et~al.}(2017)\citenamefont {Park},
  \citenamefont {Kong}, \citenamefont {Zhou},\ and\ \citenamefont
  {Cui}}]{park2017large}%
  \BibitemOpen
  \bibfield  {author} {\bibinfo {author} {\bibfnamefont {J.-H.}\ \bibnamefont
  {Park}}, \bibinfo {author} {\bibfnamefont {L.}~\bibnamefont {Kong}}, \bibinfo
  {author} {\bibfnamefont {Y.}~\bibnamefont {Zhou}},\ and\ \bibinfo {author}
  {\bibfnamefont {M.}~\bibnamefont {Cui}},\ }\href@noop {} {\bibfield
  {journal} {\bibinfo  {journal} {Nat. Methods}\ }\textbf {\bibinfo {volume}
  {14}},\ \bibinfo {pages} {581} (\bibinfo {year} {2017})}\BibitemShut
  {NoStop}%
\bibitem [{\citenamefont
  {Charbon}(2014)}]{Charbon_PTRSLA14:SinglePhotonImagingSPADReview}%
  \BibitemOpen
  \bibfield  {author} {\bibinfo {author} {\bibfnamefont {E.}~\bibnamefont
  {Charbon}},\ }\href@noop {} {\bibfield  {journal} {\bibinfo  {journal}
  {Philos. Trans. R. Soc. A}\ }\textbf {\bibinfo {volume} {372}},\ \bibinfo
  {pages} {20130100} (\bibinfo {year} {2014})}\BibitemShut {NoStop}%
\bibitem [{\citenamefont {Gariepy}\ \emph {et~al.}(2015)\citenamefont
  {Gariepy}, \citenamefont {Krstajic}, \citenamefont {Robert~Henderson},
  \citenamefont {Thomson}, \citenamefont {Buller}, \citenamefont {Heshmat},
  \citenamefont {Raskar}, \citenamefont {Leach},\ and\ \citenamefont
  {Faccio}}]{Gariepy_NatComm15:SinglePhotonFlightImaging}%
  \BibitemOpen
  \bibfield  {author} {\bibinfo {author} {\bibfnamefont {G.}~\bibnamefont
  {Gariepy}}, \bibinfo {author} {\bibfnamefont {N.}~\bibnamefont {Krstajic}},
  \bibinfo {author} {\bibfnamefont {C.~L.}\ \bibnamefont {Robert~Henderson}},
  \bibinfo {author} {\bibfnamefont {R.~R.}\ \bibnamefont {Thomson}}, \bibinfo
  {author} {\bibfnamefont {G.~S.}\ \bibnamefont {Buller}}, \bibinfo {author}
  {\bibfnamefont {B.}~\bibnamefont {Heshmat}}, \bibinfo {author} {\bibfnamefont
  {R.}~\bibnamefont {Raskar}}, \bibinfo {author} {\bibfnamefont
  {J.}~\bibnamefont {Leach}},\ and\ \bibinfo {author} {\bibfnamefont
  {D.}~\bibnamefont {Faccio}},\ }\href {https://doi.org/10.1038/ncomms7021}
  {\bibfield  {journal} {\bibinfo  {journal} {Nat. Commun.}\ }\textbf {\bibinfo
  {volume} {6}},\ \bibinfo {pages} {6021} (\bibinfo {year} {2015})}\BibitemShut
  {NoStop}%
\bibitem [{\citenamefont {Edgar}\ \emph {et~al.}(2012)\citenamefont {Edgar},
  \citenamefont {Tasca}, \citenamefont {Izdebski}, \citenamefont {Warburton},
  \citenamefont {Leach}, \citenamefont {Agnew}, \citenamefont {Buller},
  \citenamefont {Boyd},\ and\ \citenamefont
  {Padgett}}]{Edgar_NatComm12:EMCCD_entanglement}%
  \BibitemOpen
  \bibfield  {author} {\bibinfo {author} {\bibfnamefont {M.}~\bibnamefont
  {Edgar}}, \bibinfo {author} {\bibfnamefont {D.}~\bibnamefont {Tasca}},
  \bibinfo {author} {\bibfnamefont {F.}~\bibnamefont {Izdebski}}, \bibinfo
  {author} {\bibfnamefont {R.}~\bibnamefont {Warburton}}, \bibinfo {author}
  {\bibfnamefont {J.}~\bibnamefont {Leach}}, \bibinfo {author} {\bibfnamefont
  {M.}~\bibnamefont {Agnew}}, \bibinfo {author} {\bibfnamefont
  {G.}~\bibnamefont {Buller}}, \bibinfo {author} {\bibfnamefont
  {R.}~\bibnamefont {Boyd}},\ and\ \bibinfo {author} {\bibfnamefont
  {M.}~\bibnamefont {Padgett}},\ }\href {https://doi.org/10.1038/ncomms1988}
  {\bibfield  {journal} {\bibinfo  {journal} {Nat. Commun.}\ }\textbf {\bibinfo
  {volume} {3}},\ \bibinfo {pages} {984} (\bibinfo {year} {2012})}\BibitemShut
  {NoStop}%
\bibitem [{\citenamefont {Lemos}\ \emph {et~al.}(2014)\citenamefont {Lemos},
  \citenamefont {Borish}, \citenamefont {Cole}, \citenamefont {Ramelow},
  \citenamefont {Lapkiewicz},\ and\ \citenamefont
  {Zeilinger}}]{Lemos_Nat14:QIm_w_undetectedPhotons}%
  \BibitemOpen
  \bibfield  {author} {\bibinfo {author} {\bibfnamefont {G.~B.}\ \bibnamefont
  {Lemos}}, \bibinfo {author} {\bibfnamefont {V.}~\bibnamefont {Borish}},
  \bibinfo {author} {\bibfnamefont {G.~D.}\ \bibnamefont {Cole}}, \bibinfo
  {author} {\bibfnamefont {S.}~\bibnamefont {Ramelow}}, \bibinfo {author}
  {\bibfnamefont {R.}~\bibnamefont {Lapkiewicz}},\ and\ \bibinfo {author}
  {\bibfnamefont {A.}~\bibnamefont {Zeilinger}},\ }\href@noop {} {\bibfield
  {journal} {\bibinfo  {journal} {Nature}\ }\textbf {\bibinfo {volume} {512}},\
  \bibinfo {pages} {409} (\bibinfo {year} {2014})}\BibitemShut {NoStop}%
\bibitem [{\citenamefont {Tasca}\ \emph {et~al.}(2013)\citenamefont {Tasca},
  \citenamefont {Aspden}, \citenamefont {Morris}, \citenamefont {Anderson},
  \citenamefont {Boyd},\ and\ \citenamefont {Padgett}}]{Tasca_OpEx13:GI_ICCD}%
  \BibitemOpen
  \bibfield  {author} {\bibinfo {author} {\bibfnamefont {D.~S.}\ \bibnamefont
  {Tasca}}, \bibinfo {author} {\bibfnamefont {R.~S.}\ \bibnamefont {Aspden}},
  \bibinfo {author} {\bibfnamefont {P.~A.}\ \bibnamefont {Morris}}, \bibinfo
  {author} {\bibfnamefont {G.}~\bibnamefont {Anderson}}, \bibinfo {author}
  {\bibfnamefont {R.~W.}\ \bibnamefont {Boyd}},\ and\ \bibinfo {author}
  {\bibfnamefont {M.~J.}\ \bibnamefont {Padgett}},\ }\href
  {https://doi.org/10.1364/OE.21.030460} {\bibfield  {journal} {\bibinfo
  {journal} {Opt. Express}\ }\textbf {\bibinfo {volume} {21}},\ \bibinfo
  {pages} {30460} (\bibinfo {year} {2013})}\BibitemShut {NoStop}%
\bibitem [{\citenamefont {Aspden}\ \emph {et~al.}(2013)\citenamefont {Aspden},
  \citenamefont {Tasca}, \citenamefont {Boyd},\ and\ \citenamefont
  {Padgett}}]{Aspden_NJP13:EPRGI_using_ICCD}%
  \BibitemOpen
  \bibfield  {author} {\bibinfo {author} {\bibfnamefont {R.~S.}\ \bibnamefont
  {Aspden}}, \bibinfo {author} {\bibfnamefont {D.~S.}\ \bibnamefont {Tasca}},
  \bibinfo {author} {\bibfnamefont {R.~W.}\ \bibnamefont {Boyd}},\ and\
  \bibinfo {author} {\bibfnamefont {M.~J.}\ \bibnamefont {Padgett}},\ }\href
  {http://stacks.iop.org/1367-2630/15/i=7/a=073032} {\bibfield  {journal}
  {\bibinfo  {journal} {New J. Phys.}\ }\textbf {\bibinfo {volume} {15}},\
  \bibinfo {pages} {073032} (\bibinfo {year} {2013})}\BibitemShut {NoStop}%
\bibitem [{\citenamefont {Fickler}\ \emph {et~al.}(2013)\citenamefont
  {Fickler}, \citenamefont {Krenn}, \citenamefont {Lapkiewicz}, \citenamefont
  {Ramelow},\ and\ \citenamefont
  {Zeilinger}}]{Fickler_SciRep:RealTimeImagingEntanglement}%
  \BibitemOpen
  \bibfield  {author} {\bibinfo {author} {\bibfnamefont {R.}~\bibnamefont
  {Fickler}}, \bibinfo {author} {\bibfnamefont {M.}~\bibnamefont {Krenn}},
  \bibinfo {author} {\bibfnamefont {R.}~\bibnamefont {Lapkiewicz}}, \bibinfo
  {author} {\bibfnamefont {S.}~\bibnamefont {Ramelow}},\ and\ \bibinfo {author}
  {\bibfnamefont {A.}~\bibnamefont {Zeilinger}},\ }\href@noop {} {\bibfield
  {journal} {\bibinfo  {journal} {Sci. Rep.}\ }\textbf {\bibinfo {volume}
  {3}},\ \bibinfo {pages} {1914} (\bibinfo {year} {2013})}\BibitemShut
  {NoStop}%
\bibitem [{\citenamefont {Machulka}\ \emph {et~al.}(2014)\citenamefont
  {Machulka}, \citenamefont {Haderka}, \citenamefont {Pe\v{r}ina},
  \citenamefont {Lamperti}, \citenamefont {Allevi},\ and\ \citenamefont
  {Bondani}}]{Machulka_OpEx14:SpatialPropofTwinbeamCorrelation}%
  \BibitemOpen
  \bibfield  {author} {\bibinfo {author} {\bibfnamefont {R.}~\bibnamefont
  {Machulka}}, \bibinfo {author} {\bibfnamefont {O.}~\bibnamefont {Haderka}},
  \bibinfo {author} {\bibfnamefont {J.}~\bibnamefont {Pe\v{r}ina}}, \bibinfo
  {author} {\bibfnamefont {M.}~\bibnamefont {Lamperti}}, \bibinfo {author}
  {\bibfnamefont {A.}~\bibnamefont {Allevi}},\ and\ \bibinfo {author}
  {\bibfnamefont {M.}~\bibnamefont {Bondani}},\ }\href
  {https://doi.org/10.1364/OE.22.013374} {\bibfield  {journal} {\bibinfo
  {journal} {Opt. Express}\ }\textbf {\bibinfo {volume} {22}},\ \bibinfo
  {pages} {13374} (\bibinfo {year} {2014})}\BibitemShut {NoStop}%
\bibitem [{\citenamefont {Morris}\ \emph {et~al.}(2015)\citenamefont {Morris},
  \citenamefont {Aspden}, \citenamefont {Bell}, \citenamefont {Boyd},\ and\
  \citenamefont {Padgett}}]{Morris_NatComm15:ImagingSinglePhotons}%
  \BibitemOpen
  \bibfield  {author} {\bibinfo {author} {\bibfnamefont {P.~A.}\ \bibnamefont
  {Morris}}, \bibinfo {author} {\bibfnamefont {R.~S.}\ \bibnamefont {Aspden}},
  \bibinfo {author} {\bibfnamefont {J.~E.}\ \bibnamefont {Bell}}, \bibinfo
  {author} {\bibfnamefont {R.~W.}\ \bibnamefont {Boyd}},\ and\ \bibinfo
  {author} {\bibfnamefont {M.~J.}\ \bibnamefont {Padgett}},\ }\href@noop {}
  {\bibfield  {journal} {\bibinfo  {journal} {Nat. Commun.}\ }\textbf {\bibinfo
  {volume} {6}},\ \bibinfo {pages} {5931} (\bibinfo {year} {2015})}\BibitemShut
  {NoStop}%
\bibitem [{\citenamefont {Chrapkiewicz}\ \emph {et~al.}(2016)\citenamefont
  {Chrapkiewicz}, \citenamefont {Jachura}, \citenamefont {Banaszek},\ and\
  \citenamefont {Wasilewski}}]{Chrapkiewicz_NatPhoton16:ImagingSinglePhotons}%
  \BibitemOpen
  \bibfield  {author} {\bibinfo {author} {\bibfnamefont {R.}~\bibnamefont
  {Chrapkiewicz}}, \bibinfo {author} {\bibfnamefont {M.}~\bibnamefont
  {Jachura}}, \bibinfo {author} {\bibfnamefont {K.}~\bibnamefont {Banaszek}},\
  and\ \bibinfo {author} {\bibfnamefont {W.}~\bibnamefont {Wasilewski}},\
  }\href@noop {} {\bibfield  {journal} {\bibinfo  {journal} {Nat. Photon.}\
  }\textbf {\bibinfo {volume} {10}},\ \bibinfo {pages} {576} (\bibinfo {year}
  {2016})}\BibitemShut {NoStop}%
\bibitem [{\citenamefont {Larson}\ and\ \citenamefont
  {Saleh}(2018)}]{larson2018resurgence}%
  \BibitemOpen
  \bibfield  {author} {\bibinfo {author} {\bibfnamefont {W.}~\bibnamefont
  {Larson}}\ and\ \bibinfo {author} {\bibfnamefont {B.~E.}\ \bibnamefont
  {Saleh}},\ }\href@noop {} {\bibfield  {journal} {\bibinfo  {journal}
  {Optica}\ }\textbf {\bibinfo {volume} {5}},\ \bibinfo {pages} {1382}
  (\bibinfo {year} {2018})}\BibitemShut {NoStop}%
\end{thebibliography}


\end{document}